\documentclass{article}
\usepackage[margin=1in]{geometry}
\usepackage{blindtext}
\usepackage{booktabs}

\usepackage[numbers]{natbib}
\usepackage{longtable} 
\usepackage{graphicx}
\usepackage{gensymb}
\usepackage{amsmath}

\usepackage{chemfig}
\usepackage{subcaption}
\usepackage{amssymb}
\usepackage{siunitx}
\usepackage{float}
\DeclareSIUnit{\atm}{atm}
\DeclareSIUnit{\hPa}{hPa}
\usepackage[version=4]{mhchem}
\usepackage{authblk}
\usepackage{hyperref}

\title{A cometary Fluorescence Model for the $\nu_{3}$ Vibrational Band of Cyanogen}
\author[1]{P. Hardy}
\author[2]{P. Rousselot}
\author[1]{C. Richard}
\author[1]{V. Boudon}
\author[3]{X. Landsheere}
\author[3,4]{A. Voute}
\author[3,4]{L. Manceron}
\author[3]{F. Kwabia Tchana}

\affil[1]{Université Bourgogne Europe, CNRS, Laboratoire Interdisciplinaire Carnot de Bourgogne ICB UMR 6303, 21000 Dijon, France}
\affil[2]{Université Marie et Louis Pasteur, CNRS, Institut UTINAM (UMR 6213), OSU THETA, F-25000 Besançon, France}
\affil[3]{Université Paris Cité and Université Paris Est Créteil, CNRS, LISA, F-75013 Paris, France}
\affil[4]{Synchrotron SOLEIL, Ligne AILES, L'Orme des Merisiers, St-Aubin BP48, 91192 Gif-sur-Yvette Cedex, France}

\begin{document}
\maketitle

\begin{abstract}

Cyanogen (\ce{C_2N_2}) has been suspected for a long time to be present in comets and to contribute to the creation of the CN radical.  So far no observations with ground-based facilities have managed to detect this species but the  Rosetta mission, thanks to in situ observations with the ROSINA mass spectrometer detected this species in the coma of 67P/Churyumov-Gerasimenko. To investigate its presence from infrared spectra in other comets, we developed a fluorescence model for the $\nu_3$ fundamental band. From new laboratory high-resolution infrared spectra of cyanogen, we analyzed the region of the $\nu_3$ band of \ce{C_2N_2}, centered around \SI{4.63}{\micro\meter} (\SI{2158}{\per\centi\meter}). In addition to line positions and intensities, molecular parameters for the ground and excited vibrational state were obtained. These parameters allowed us to develop a fluorescence model for cyanogen. Excitation rates of the $\nu_3$ band of cyanogen in cometary comae are presented. An attempt to detect cyanogen in a high-resolution spectrum of comet C/2022 E3 (ZTF) is discussed.

\end{abstract}

\section{Introduction}
Among the limited number of space missions dedicated to cometary science, the Rosetta spacecraft launched in 2004 significantly enhanced the understanding of those small bodies, by conducting a two-year study of the comet 67P/Churyumov-Gerasimenko. One of Rosetta’s instruments, the mass spectrometer ROSINA, discovered numerous chemical species in the coma that had never been observed in comets before.\\
One such species, cyanogen (\ce{C_2N_2}) is a linear molecule that was previously only observed in the atmosphere of Titan by \cite{Kunde_1981} before its detection on comet 67P \citep{Altwegg_Balsiger_Fuselier_2019}. Although the CN radical is a prominent feature in cometary optical spectra, its origins (i.e., its corresponding parent species) are still poorly understood. In particular, hydrogen cyanide (HCN) has been shown not to be the only source of cometary CN \citep{Hanni_Altwegg_Pestoni_Rubin_Schroeder_Schuhmann_Wampfler_2020,Hanni_Altwegg_Balsiger_Combi_Fuselier_De_Keyser_Pestoni_Rubin_Wampfler_2021}. \ce{C_2N_2}, while being much less abundant than HCN in the coma of comet 67P (\cite{Hanni_Altwegg_Balsiger_Combi_Fuselier_De_Keyser_Pestoni_Rubin_Wampfler_2021} reported $\ce{C_2N_2}/\ce{HCN}=0.0018\pm 0.0009$, (1.24-1.74 au)) could still participate in the formation of CN. Its detection in 67P should be extended to other comets to better constrain its abundance in cometary environments.\\
While the $\nu_5$ far-infrared band of \ce{C_2N_2} is already present in the HITRAN database \citep{hitran20,fayt}, no data are available at wavenumbers greater than \SI{310}{\per\centi\meter}. In particular, one of the fundamental bands of cyanogen, centered at \SI{2158}{\per\centi\meter} (\SI{4.63}{\micro\meter}) was never studied at high-resolution. This band is particularly interesting to study as it is centered in the M-infrared window, a spectral region where few atmospheric emission lines of \ce{H_2O} and \ce{CO_2} are present. \\
The paper is organized as follows. After discussing experimental details (Section \ref{sec:expe}) and theoretical aspects (Section \ref{sec:theory}), we present both the results of our spectroscopic analysis in Section \ref{sec:spectro} and the development of a fluorescence model adapted to comets in Section \ref{sec:fluo}. Finally, our results and an upper limit on the abundance of cyanogen in comet C/2022 E3 (ZTF) is presented in Section \ref{sec:results}.

\section{Experimental details}\label{sec:expe}
Seven absorption spectra of cyanogen have been recorded in the range from 1800 to 3500 \SI{}{\per\centi\meter} using the high-resolution Bruker IFS125HR Fourier transform spectrometer (FTS) located at the LISA facility in Créteil (France). The instrument was equipped with a silicon carbide Globar source, a KBr/Ge beamsplitter and a liquid nitrogen-cooled indium antimonide (InSb) detector. The InSb detector was used in conjunction with an optical filter, with a bandpass of 1800–3400~\SI{}{\per\centi\meter}, to minimize the size of the interferogram data files and also to improve the signal-to-noise ratio. The FTS was continuously evacuated below $3\times 10^4$ hPa by a turbomolecular pump to minimize absorption by atmospheric gases. The diameter of the entrance aperture of the spectrometer was set to 1.3 mm to maximize the intensity of infrared radiation falling onto the InSb detector without saturation or loss of spectral resolution. Interferograms were recorded with an 80 kHz scanner frequency and a maximum optical path difference (MOPD) of 250 cm. According to the Bruker definition (resolution = 0.9/MOPD), this corresponds to a resolution of 0.0036 \SI{}{\per\centi\meter}. The spectra were obtained by Fourier transformation of the interferograms using a Mertz phase correction with a \SI{1}{\per\centi\meter} resolution, a zero-filling factor of 2 and no apodization (boxcar option).
The cyanogen (\ce{C_2N_2}) sample was synthesized in the laboratory by progressive heating of silver cyanide (99\%, Sigma Aldrich) in a glass bulb previously evacuated:
\begin{equation}
\schemestart
2 \chemfig[atom sep=2em]{Ag-CN}
\arrow{->[$\Delta > 350^{\circ}\text{C}$][{(-2Ag)}]}
\chemfig[atom sep=2em]{NC-CN}.
\schemestop
\end{equation}

About 5 g of silver cyanide powder was placed in a thoroughly cleaned, grease-free glass bulb and heated gently under high vacuum up to 200\degree C to eliminate atmospheric gases and adsorbed water vapor. Next, the powder was progressively heated to 300\degree C where evolution of cyanogen gas slowly started to occur. The first batch, containing traces of carbon dioxide from carbonate impurities, was discarded, further heating to about 400\degree C completed the reaction and cyanogen was collected in a glass trap on the vacuum line. Finally, a trap-to-trap distillation at about -70\degree C helped reduce possible remaining impurities (\ce{HCN}, \ce{CO_2} or others). The remaining \ce{CO_2} traces were estimated from IR measurements to much less than 0.1\%.

A short-path absorption cell (SPAC) made of Pyrex glass and equipped with CsBr windows was used for all the measurements. The SPAC is a White-type multipass cell with a base length of 0.20 m. In this experiment, an optical path of 0.849(2) m was used. The sample pressure in the cell was measured using calibrated MKS Baratron capacitance manometers models 627 (2.6664 hPa full scale) and 628 D (13.332 hPa full scale) characterized by a stated reading accuracy of 0.12\%. Considering the uncertainty arising from small variations of the pressure during the recording ($< 0.35\%$), we estimated the measurement uncertainty on the pressure to be equal to 0.5\%. All the spectra were recorded at a stabilized room temperature of $293 \pm 1$ K.
The following procedure was used to record the spectra. A background spectrum was first collected while the cell was being continuously evacuated. It was recorded at the same resolution as the sample spectra to ensure proper removal of the \ce{H_2O} or \ce{CO_2} absorption lines. The infrared gas cell was then passivated several times with the \ce{C_2N_2} sample. Finally, spectra were recorded for seven different sample pressures of cyanogen. The seven pressures chosen and the number of interferograms recorded and averaged to yield the corresponding spectra are listed in Table~\ref{tab:expe}. All the sample spectra were ratioed against the empty cell background spectrum, and interpolated 4 times. The rms signal-to-noise ratio in the ratioed spectra ranged between 1200 and 1900. The spectrum was calibrated by matching the measured positions of 25 lines of residual \ce{CO_2} and \ce{H_2O} observed therein to reference wavenumbers available in HITRAN (\cite{hitran20}) with an rms deviation of 0.00018 \SI{}{\per\centi\meter}. The absolute accuracy of the measured \ce{C_2N_2} line positions was estimated as the square root of the sum of squares of the accuracy of the reference \ce{CO_2} and \ce{H_2O} line positions in HITRAN (better than \SI{0.0001}{\per\centi\meter}) plus the RMS deviation between the observed frequencies and values listed in HITRAN for \ce{CO_2} and \ce{H_2O}, yielding in total \SI{0.0002}{\per\centi\meter}.

\begin{table}[]
\centering
\begin{tabular}{lll}
\hline
\# & \ce{C_2N_2} Pressure (hPa)    & Number of Scans \\ \hline
S1 & 0.9968 (50) & 1460     \\
S2 & 1.3404 (67) & 1310     \\
S3 & 1.8718 (94) & 1430     \\
S4 & 2.337 (12)  & 1410     \\
S5 & 2.673 (13)  & 1480     \\
S6 & 3.335 (17)  & 1250     \\
S7 & 4.003 (20)  & 1250     \\ \hline
\end{tabular}
\caption{Pressure of \ce{C_2N_2} (in hPa) and number of interferograms averaged to yield the corresponding spectrum (number of scans). All the spectra were recorded with an absorption optical path of 0.849(2) m, at a stabilized room temperature of $293 \pm 1$ K, a resolution (equal to 0.9 divided by the maximum optical path difference) of \SI{0.0036}{\per\centi\meter} and an entrance aperture diameter of the interferometer equal to 1.3 mm. The absolute uncertainty on the pressure is equal to 0.5\% of the value given.}
\label{tab:expe}
\end{table}
\section{Theoretical aspects}\label{sec:theory}
Cyanogen (\ce{C2N2}) is a linear molecule belonging to the $D_{\infty h}$ point group, with five fundamental vibrational modes: CN symmetric stretch ($\nu_1$, centered at \SI{2330.4859}{\per\centi\meter}, with $\Sigma_g$ symmetry), CC symmetric stretch ($\nu_2$, centered at \SI{845.5939}{\per\centi\meter}, with $\Sigma_g$ symmetry, infrared inactive), CN asymmetric stretch ($\nu_3$, centered at \SI{2157.8243}{\per\centi\meter}, with $\Sigma_u$ symmetry), and two doubly degenerated CCN bending modes ($\nu_4$, centered at \SI{502.7745}{\per\centi\meter}, with $\Pi_g$ symmetry, infrared inactive; and $\nu_5$, centered at \SI{233.7225}{\per\centi\meter}, with $\Pi_u$ symmetry). Band-center values come from the analysis of \cite{MAKI2011166}. Our study is focused on the $\nu_3$ mode, centered at 2158 cm$^{-1}$.

The rotational energy levels are described as usual by
\begin{equation}
E(J)  = B J (J+1) - D J^2 (J+1)^2 + H J^3 (J+1)^3+\dots,
\end{equation}
where $B$ is the rotational constant, and $D$ and $H$ represent centrifugal distortion corrections. The allowed frequencies are obtained as $\nu_J = \nu_0 + E_{u}(J_{u}) - E_{l}(J_{l})$, where $\nu_0$ is the frequency of the pure vibrational transitions, and the subscripts $u$ and $l$ refer to the upper and lower vibrational levels, respectively. For non-$\Sigma$ states, the interaction between rotation and angular momentum components leads to $l$-doubling, resulting in energy splitting given by
\begin{equation}
\Lambda(J)=qJ(J+1)+q_DJ^2(J+1)^2+q_HJ^3(J+1)^3+\dots
\end{equation}

Individual line strengths ($S_J$) can be expressed in terms of the band strength ($S_\nu$, usually given in \SI{}{\per\centi\meter\squared\per\atm}) \citep{rothman}
\begin{equation}\label{eq:int}
    S_J = \varepsilon_J \frac{\nu_J}{\nu_0}L_J\frac{S_\nu}{Q_r}(e^{-hcE_l/kT})(1-e^{-hc\nu_J/kT})(1+A_1m)^2.
\end{equation}
Here, $\varepsilon_J$ represents the statistical weight, $L_J$ is the Hönl-London factor, and $Q_R$ is the rotational partition function. The last term is the first-order Herman-Wallis correction, which is equal to 1 for perfect rigid rotors and $Q$ branches. This correction depends on $m$, defined as $J+1$ for the $R$ branch, and $-J$ for the $P$ branch. For linear (and diatomic) molecules, the Hönl-London factors for the $P$ and $R$ branches of $\Sigma-\Sigma$ transitions are simply given by \citep{herzbergvol1}
\begin{equation}\label{eq:br}
    L_J= |m|.
\end{equation}
For the main isotopologue (\ce{^{14}N^{12}C^{12}C^{14}N}, with an abundance of 97.08\% reported in \cite{hitran20}), $\varepsilon_J$ is equal to 6 for even J values; 3 for odd J values. In practice, these statistical weights lead to an intensity alternation between two consecutive lines in the spectra. The rotational partition function appearing in equation \ref{eq:int} is expressed as
\begin{equation}
    Q_R=\sum_{J=0}^{\infty} \varepsilon_J (2J+1) e^{-hcE_{l}(J)/kT},
\end{equation}
while the vibrational partition function is given by
\begin{equation}
    Q_V=\prod_{i=1}^{5} \left(1-e^{(-hc\nu_{0,i}/kT)}\right)^{-d_i}.
\end{equation}
where $d_i$ is the degeneracy of the fundamental band $i$, centered at $\nu_{0,i}$. The rotovibrational partition function ($Q$) is the product of $Q_R$ and $Q_V$.
To derive the excitation rates, we followed the development presented in \cite{RUSSO2001162}. 
The Einstein coefficient $A_\nu$ (in \SI{}{\per\second}) for a vibrational band \citep{crovisier_encrenaz_1983,crovisier_1984} is defined by :
\begin{equation}\label{eq:einstein}
    A_\nu=\frac{3.080 \times 10^{-8} (T_\nu/300)\nu_0^{2}S_{\nu}Q_V}{\omega},
\end{equation}
where $T_\nu$ is the temperature (in kelvins) at which the band strength $S_\nu$ (in \SI{}{\per\centi\meter\squared\per\atm}) is obtained, and $\omega$ is the band degeneracy ($\omega=1$ for $\nu_3$). The constant $3.080\ \times 10^{-8}$ The $g$-factor for a fundamental band excited by a blackbody of solid angle $\Omega$ at temperature $T_b$ is then expressed \citep{crovisier_encrenaz_1983}:
\begin{equation}\label{eq:gnu}
    g_\nu=\frac{\omega\Omega}{4\pi}A_\nu\left(e^{hc\nu_0/kT_b}-1\right)^{-1}.
\end{equation}
With an excitation by the Sun at 1 au, $T_b$=5770 K and $\Omega/4\pi=5.42\times10^{-6}$ \citep{RUSSO2001162}.

\section{Spectroscopic analysis}\label{sec:spectro}
\subsection{Positions}
\begin{figure*}
    \centering
    \includegraphics[width=0.8\linewidth]{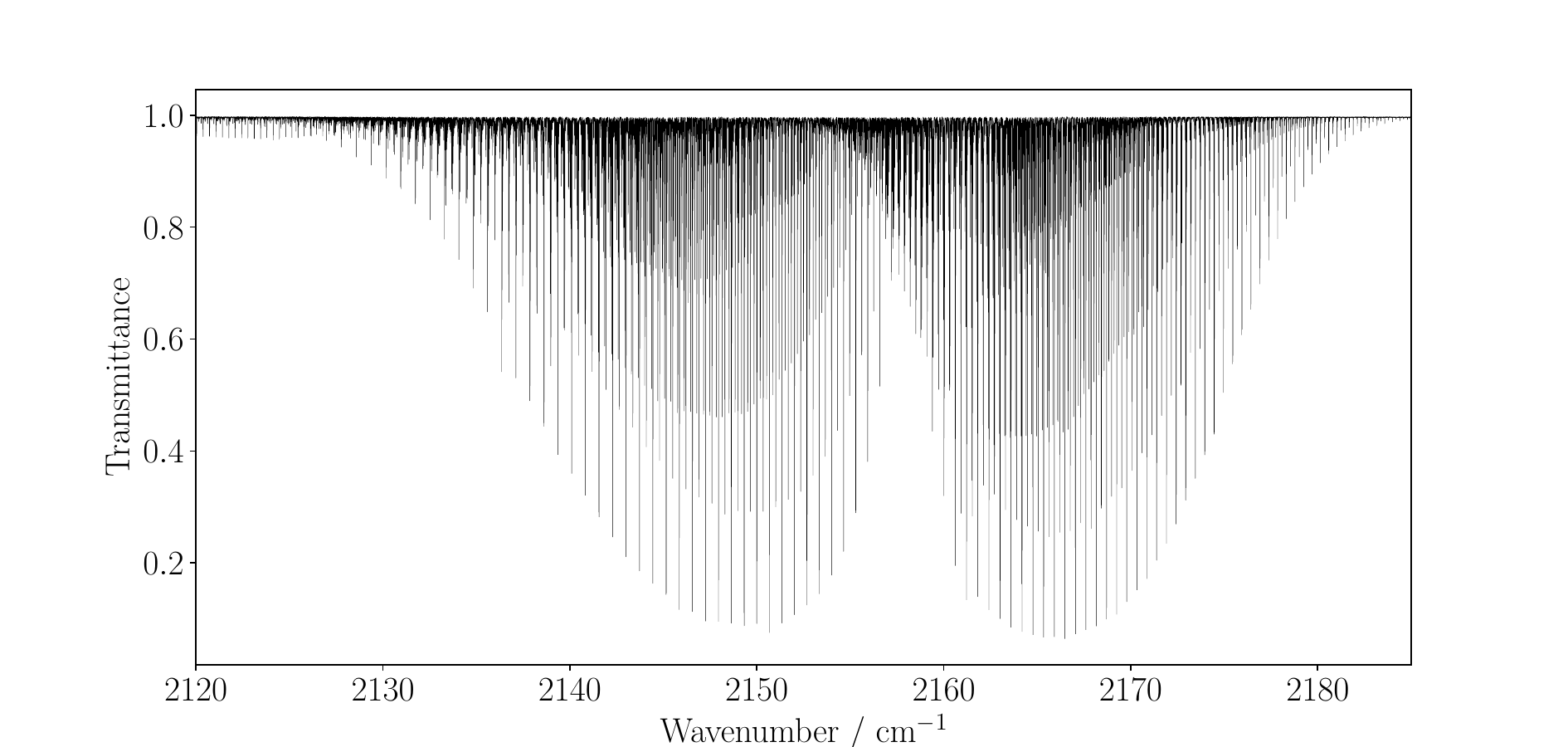}
    \caption{Experimental high-resolution spectrum of \ce{C_2N_2} between 2120 and 2185 \SI{}{\per\centi\meter}}
    \label{fig:spectrum}
\end{figure*}

\begin{figure}
    \centering
    \includegraphics[width=0.4\linewidth]{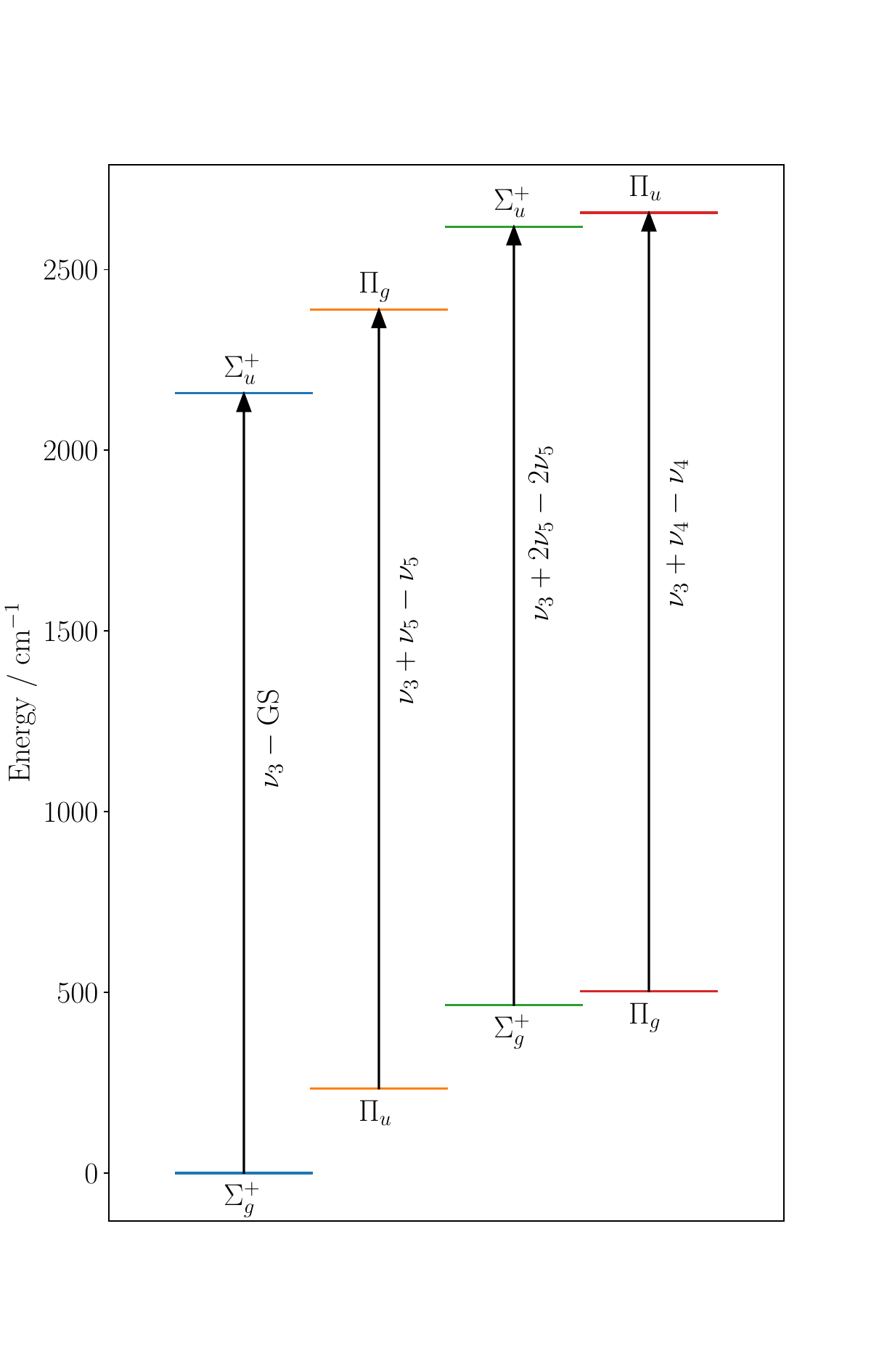}
    \caption{The four vibrational bands presented in this work: $\nu_3$, $\nu_3+\nu_5-\nu_5$, $\nu_3+2\nu_5-2\nu_5$, and $\nu_3+\nu_4-\nu_4$. Symmetries are labeled for each energy level.}
    \label{fig:bands}
\end{figure}

Using the PGOPHER software \citep{WESTERN2017221}, we analyzed the experimental spectra of \ce{C_2N_2} from 2120 \SI{}{\per\centi\meter} to 2185 \SI{}{\per\centi\meter}. The S2 experimental spectrum (P=1.3404(67) hPa) is represented in Fig. \ref{fig:spectrum}.
In addition to the $\nu_3$ fundamental band, some absorption lines from hot bands are also present in the spectrum: we identified $\nu_3+\nu_4-\nu_4$, $\nu_3+\nu_5-\nu_5$, and $\nu_3+2\nu_5-2\nu_5$. Energy levels and the four transitions discussed hereafter are shown in Fig. \ref{fig:bands}.
As $\nu_4$, $\nu_5$ and $2\nu_5$ are the lowest excited energy levels of $\ce{C_2N_2}$, these hot bands are quite strong and affect the study of the fundamental band $\nu_3$, because hot bands lines are often blended with $\nu_3$ lines. In this context and because they were fairly easy to assign, we decided to include the hot bands in this work, even though the fluorescence model described later only focuses on the fundamental band, as it is supposed to be the main component in cold environments such as the coma of a comet.  We started our analysis in PGOPHER using spectroscopic parameters from \cite{MAKI2011166}.

For the $\nu_3$ band, 201 lines have been assigned from $J=0$ to $J=112$. A few lines have been systematically removed from the fit if they were obviously blended with other lines, and decreased the quality of the fit. The average error $d_\mathrm{RMS}$, defined as
\begin{equation}
    d_{RMS}=\sqrt{\frac{\sum_i d_i^2}{n}},
\end{equation}
where $n$ is the number of observations, and $d_i$ is the difference in \SI{}{\per\centi\meter} between the observed and calculated positions of line $i$. In our fit of the fundamental band $\nu_3$, $d_\mathrm{RMS}=1.21 \times 10^{-4}~\mathrm{cm^{-1}}$. For the hot band $\nu_3+2\nu_5-2\nu_5$, 114 lines have been assigned, from $J=1$ to $J=99$, with  $d_{RMS}=2.43\times 10^{-4}~\mathrm{cm^{-1}}$. $\nu_3$ and $\nu_3+2\nu_5$ are both $\Sigma_u^+$ states, meaning there is no $\Lambda$-doubling observed for those two bands. This is not the case for $\nu_3+\nu_4$ and $\nu_3+\nu_5$, which are respectively $\Pi_u$ and $\Pi_g$ states. As lines are blended for low $J$ (because the splitting is not big enough), we did not include lines with $J<11$ for $\nu_3+\nu_5-\nu_5$ and $J<14$ for $\nu_3+\nu_4-\nu_4$ while fitting positions. The evolution of the splitting as $J$ increases is illustrated in Fig. \ref{fig:lambda}.

\begin{figure*}
    \centering
    \begin{subfigure}{0.45\textwidth}
        \centering
        \includegraphics[width=\textwidth]{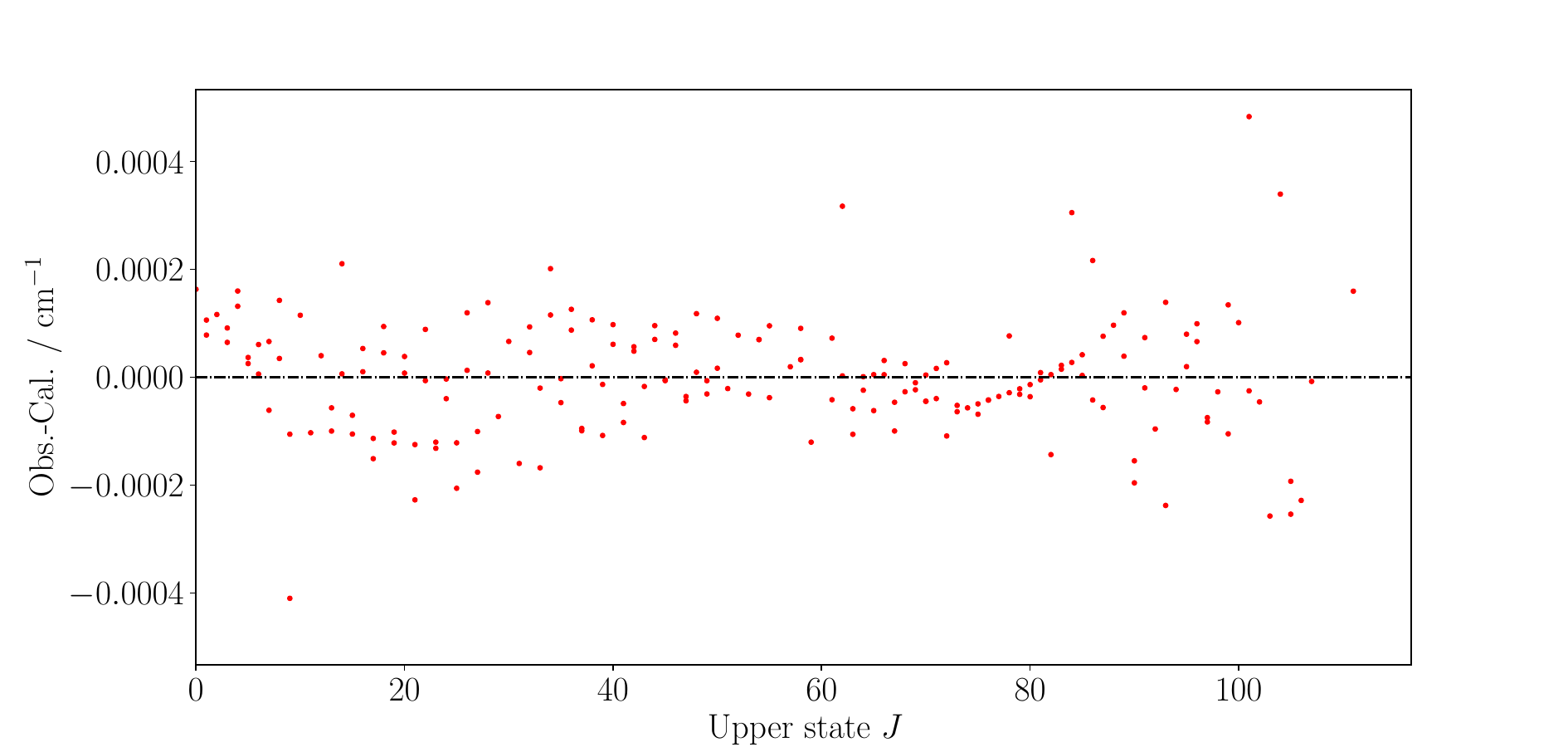}
        \caption{Fit residuals for the $\nu_3$ band. 190 observations, average error: $1.10\times10^{-4}$~$\mathrm{cm^{-1}}$.}
        \label{fig:resnu3}
    \end{subfigure}
    \hfill
    \begin{subfigure}{0.45\textwidth}
        \centering
        \includegraphics[width=\textwidth]{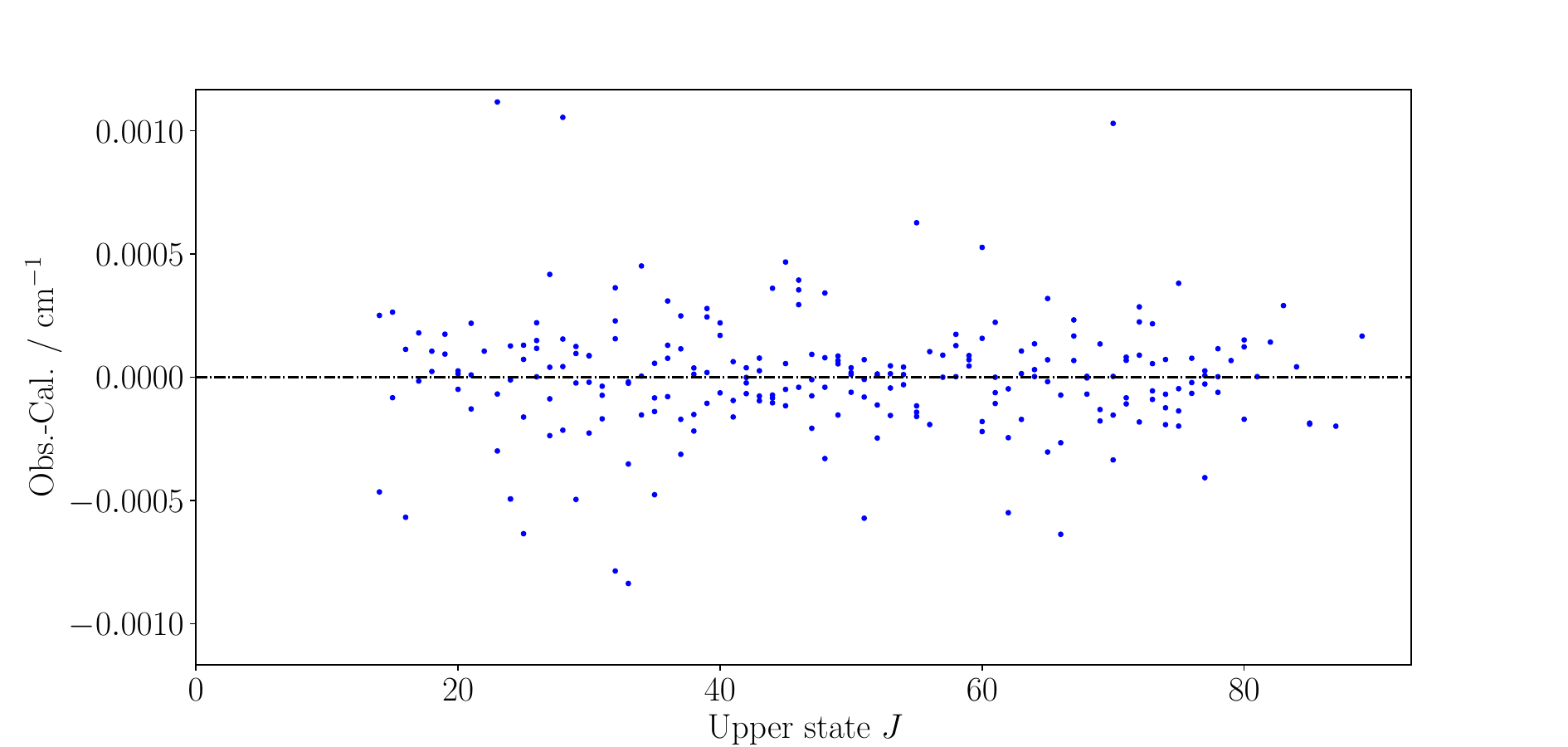}
        \caption{Fit residuals for the $\nu_3+\nu_4-\nu_4$ band. 234 observations, average error: $2.44\times10^{-4}$~$\mathrm{cm^{-1}}$.}
        \label{fig:resnu3nu4}
    \end{subfigure}
    
    \vspace{0.5cm}
    
    \begin{subfigure}{0.45\textwidth}
        \centering
        \includegraphics[width=\textwidth]{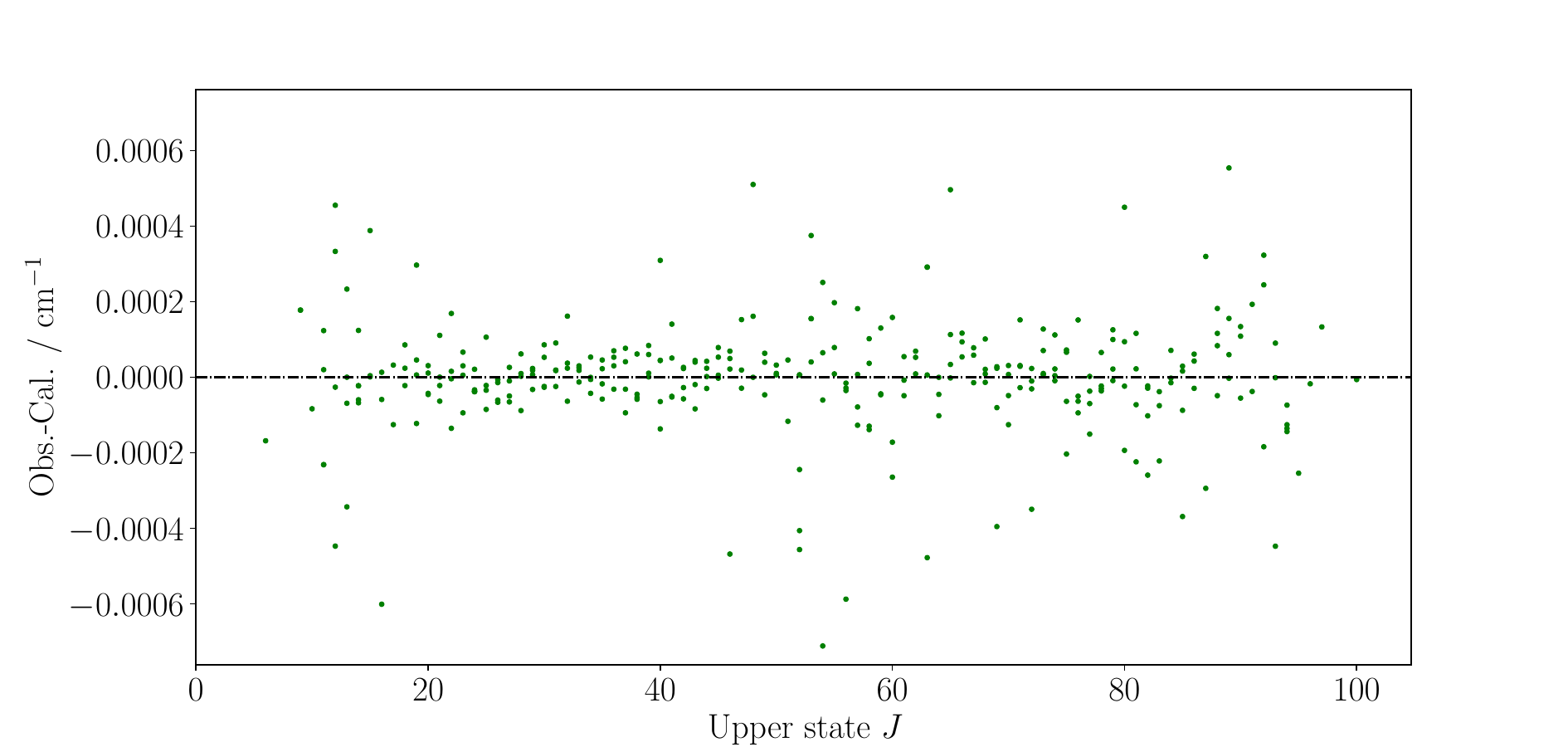}
        \caption{Fit residuals for the $\nu_3+\nu_5-\nu_5$ band. 319 observations, average error: $1.53\times10^{-4}$~$\mathrm{cm^{-1}}$.}
        \label{fig:resnu3nu5}
    \end{subfigure}
    \hfill
    \begin{subfigure}{0.45\textwidth}
        \centering
        \includegraphics[width=\textwidth]{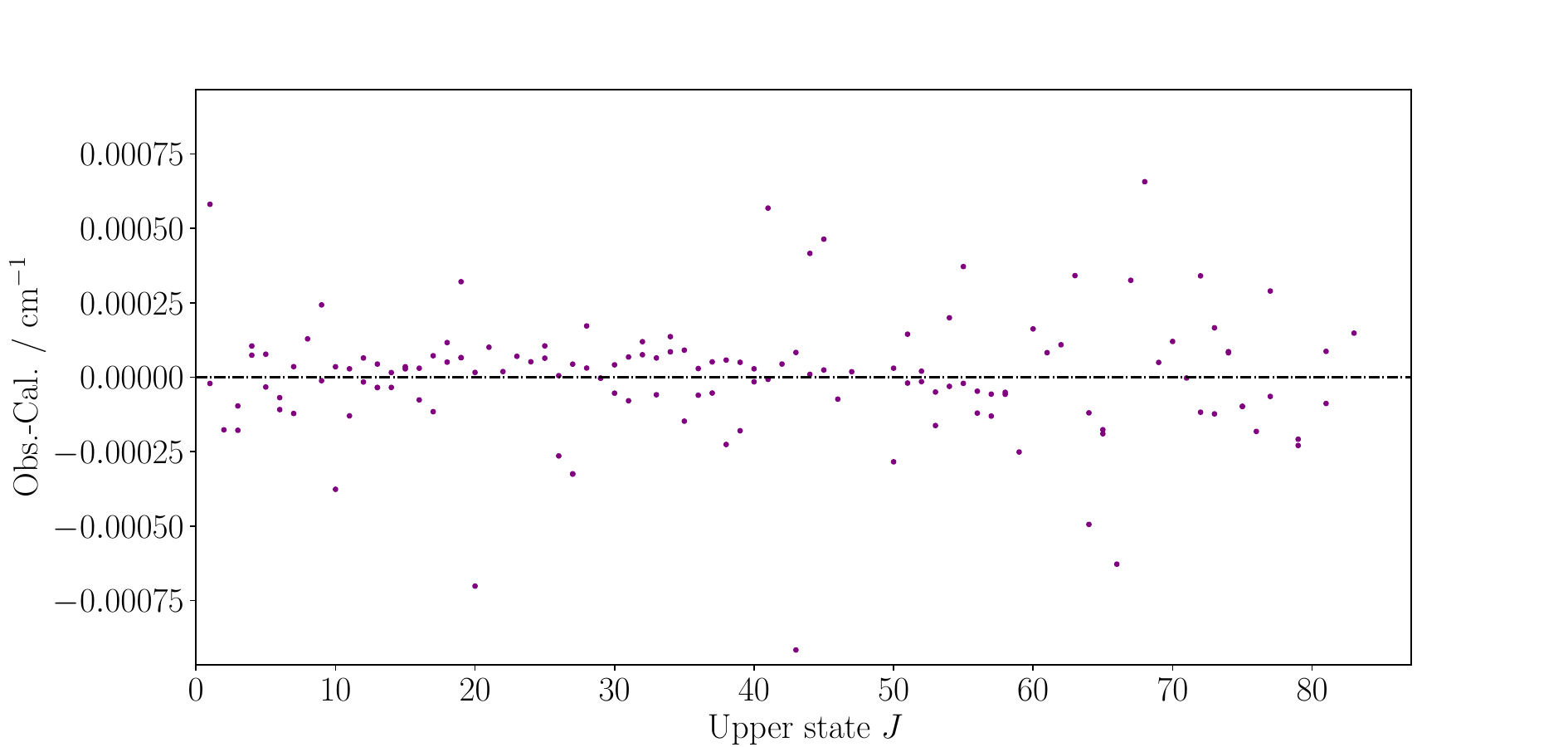}
        \caption{Fit residuals for the $\nu_3+2\nu_5-2\nu_5$ band. 137 observations, average error: $2.06\times10^{-4}$~$\mathrm{cm^{-1}}$.}
        \label{fig:resnu32nu5}
    \end{subfigure}
    
    \vspace{0.5cm}
    
    \begin{subfigure}{0.45\textwidth}
        \centering
        \includegraphics[width=\textwidth]{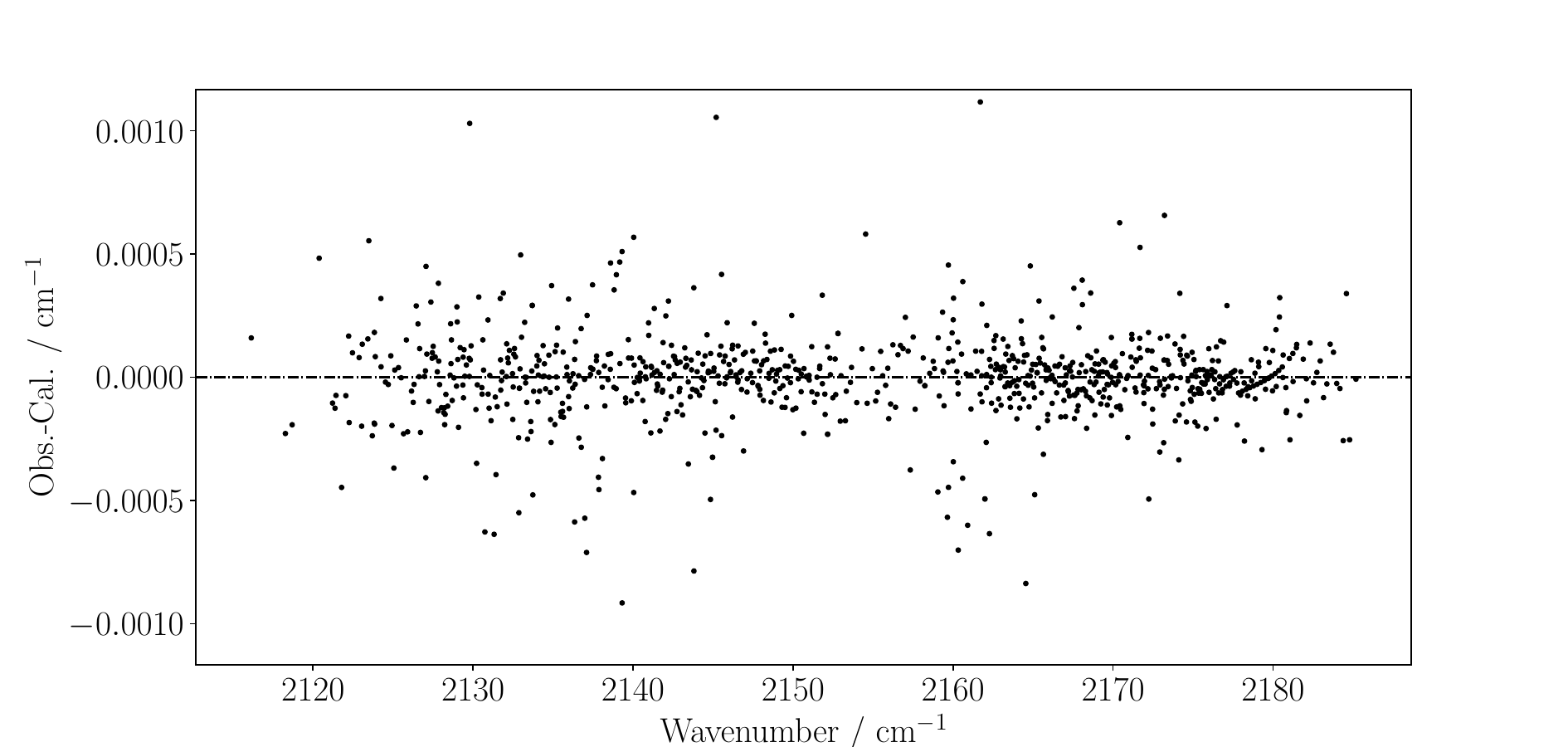}
        \caption{Fit residuals for all four bands, plotted against wavenumber. 880 observations, average error: $1.83\times10^{-4}$~$\mathrm{cm^{-1}}$.}
        \label{fig:res}
    \end{subfigure}

    \caption{Fit residuals for the line positions of the vibrational bands of $\ce{C_2N_2}$ with respect to the upper-state rotational quantum number $J$ and wavenumber.}
    \label{fig:combined_residuals}
\end{figure*}

For $\nu_3+\nu_4-\nu_4$, we fitted 195 lines from $J=13$ to $J=85$, with $d_{RMS}=2.34\times 10^{-4}~\mathrm{cm^{-1}}$. For $\nu_3+\nu_5-\nu_5$, we fitted 323 lines from $J=10$ to $J=96$, with $d_{RMS}=1.46\times 10^{-4}~\mathrm{cm^{-1}}$. The fit residuals for the line positions of each band discussed are represented in Fig. \ref{fig:resnu3}, \ref{fig:resnu3nu4}, \ref{fig:resnu3nu5}, and \ref{fig:resnu32nu5}. Residuals are plotted against the upper-level rotational quantum number $J$, to highlight potential nonzero higher-terms in the polynomial expansion of the energy states. The total average error (considering the four bands together) is $1.8\times 10^{-4}~\mathrm{cm^{-1}}$, and all the fit residuals are represented in Fig. \ref{fig:res}, where the $x$-axis is now the wavenumber in $\mathrm{cm^{-1}}$.

The spectroscopic parameters derived from the fit in positions are summarized in Table \ref{tab:para}. The standard deviation is indicated in parentheses. No parentheses indicate that the value was fixed using values from \cite{MAKI2011166}.
\begin{table*}[]
\resizebox{\textwidth}{!}{
\begin{tabular}{llllllllll}
\toprule
 & Origin & B & D $\times\ 10^{8}$ & H $\times\ 10^{13}$ & L $\times\ 10^{17}$ & q $\times\ 10^{4}$& qD $\times\ 10^{9}$& qH $\times\ 10^{13}$ & qL $\times\ 10^{17}$ \\
\midrule
GS $(\Sigma_g)$ & 0 & 0.1570879982(2801) & $2.11469(189)$ & 0& 0& 0& 0& 0& 0 \\
$\nu_5$ $(\Pi_u)$ & $233.7225^{a}$ & 0.1576278497(3905) & $2.222624(8402)$ & $0.11146(5439)$& 0& $-2.212580(4536)$& $0.4129(387)$& 0& 0 \\
$2\nu_5$ $(\Sigma_g)$ & $464.8685^{a}$ & 0.158166967(1048) & $4.20455(5699)$ & $9.153(1175)$& $-1.7012(8007)$& 0& 0& 0& 0 \\
$\nu_4$ $(\Pi_g)$ & $502.7745^{a}$ & 0.1574531178(2931) & $2.15394(332)$ & 0& 0& $-1.17900(1700)$& $-3.3476(8270)$& $5.628(1551)$& $-3.3370(9800)$ \\
$\nu_3$ $(\Sigma_u)$ & 2157.824256(34) & 0.1565544657(2892) & $2.101485(3062)$ & $-0.06649(2975)$& $0.0300(132)$& 0& 0& 0& 0 \\
$\nu_3+\nu_5$ $(\Pi_g)$ & 2389.736280(30) & 0.1570968635(3867) & $2.214595(8253)$ & $0.09869(5294)$& 0& $-2.186458(4525)$& $0.3904(383)$& 0& 0 \\
$\nu_3+2\nu_5$ $(\Sigma_u)$ & 2619.092571(39) & 0.157638754(1055) & $4.17941(5716)$ & $9.163(1170)$& $-1.8518(7922)$& 0& 0& 0& 0 \\
$\nu_3+\nu_4$ $(\Pi_u)$ & 2657.530013(67) & 0.1569204218(3257) & $2.160948(8013)$ & $0.3876(1410)$& $-0.3061(958)$& $-1.17655(1709)$& $-3.2873(8441)$& $5.4587(16123)$& $-3.192(1040)$ \\
\bottomrule
\end{tabular}

}
\caption{Derived parameters from the line positions analysis in PGOPHER. GS stands for Ground State. Uncertainties are indicated in parentheses, in units of the last digits. $B$: rotational constant, $D$: $J^4$ centrifugal distortion, $H$: $J^6$ centrifugal distortion, $L$: $J^8$ centrifugal distortion, $q$: lambda-doubling constant, $qD$: $J^4$ centrifugal distortion of $q$, $qH$: $J^6$ centrifugal distortion of $q$, $qL$: $J^8$ centrifugal distortion of $q$. $^{a}$fixed values coming from \cite{MAKI2011166}, all values are in units of $\mathrm{cm^{-1}}$.}
\label{tab:para}
\end{table*}

\clearpage
\subsection{Intensities}
\begin{figure*}
    \centering
    \begin{minipage}{\textwidth}
        \centering
        \includegraphics[width=0.5\textwidth]{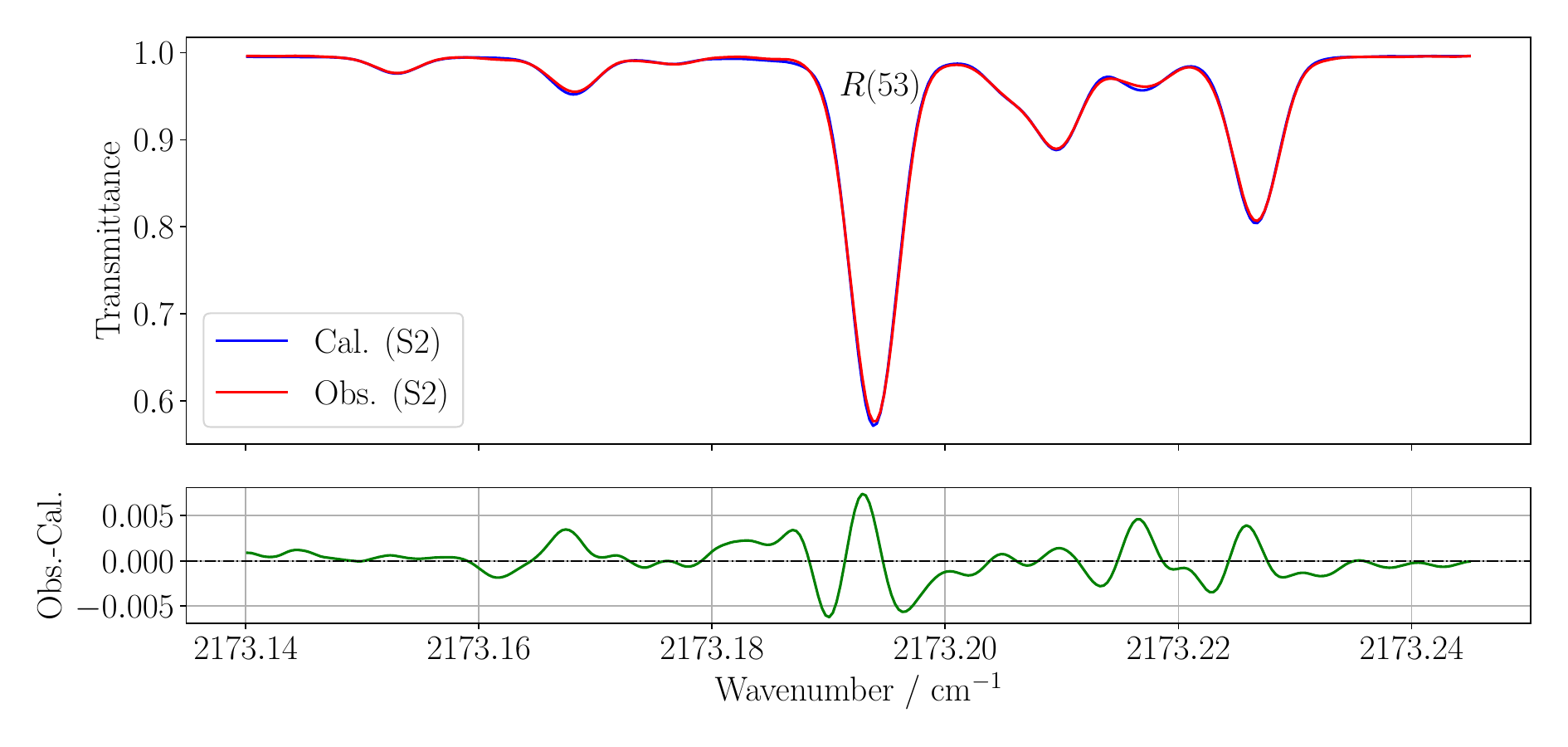}
        \subcaption{}
    \end{minipage}
    
    \vspace{0.2em} 
    
    \begin{minipage}{\textwidth}
        \centering
        \includegraphics[width=0.5\textwidth]{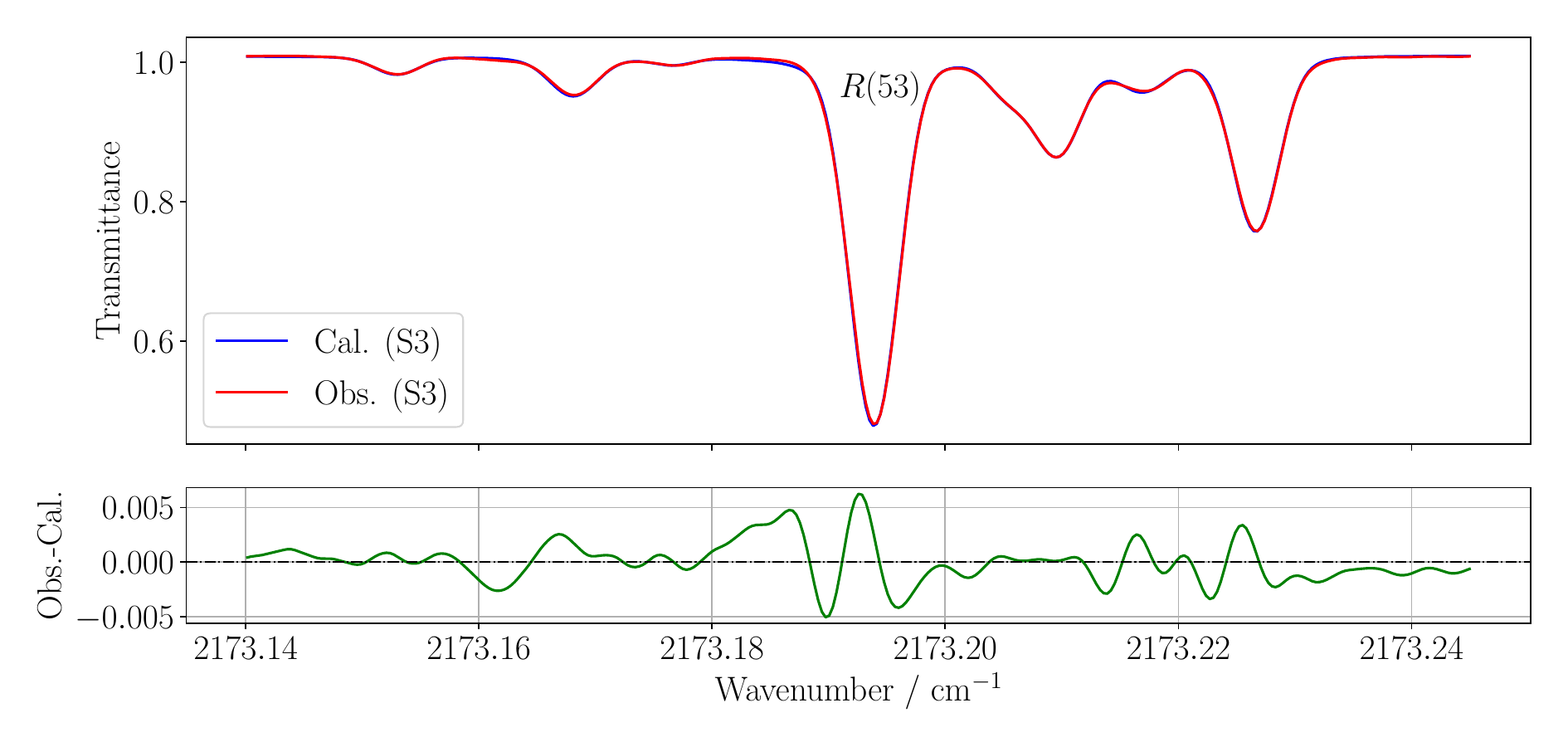}
        \subcaption{}
    \end{minipage}
    
    \vspace{0.2em}
    
    \begin{minipage}{\textwidth}
        \centering
        \includegraphics[width=0.5\textwidth]{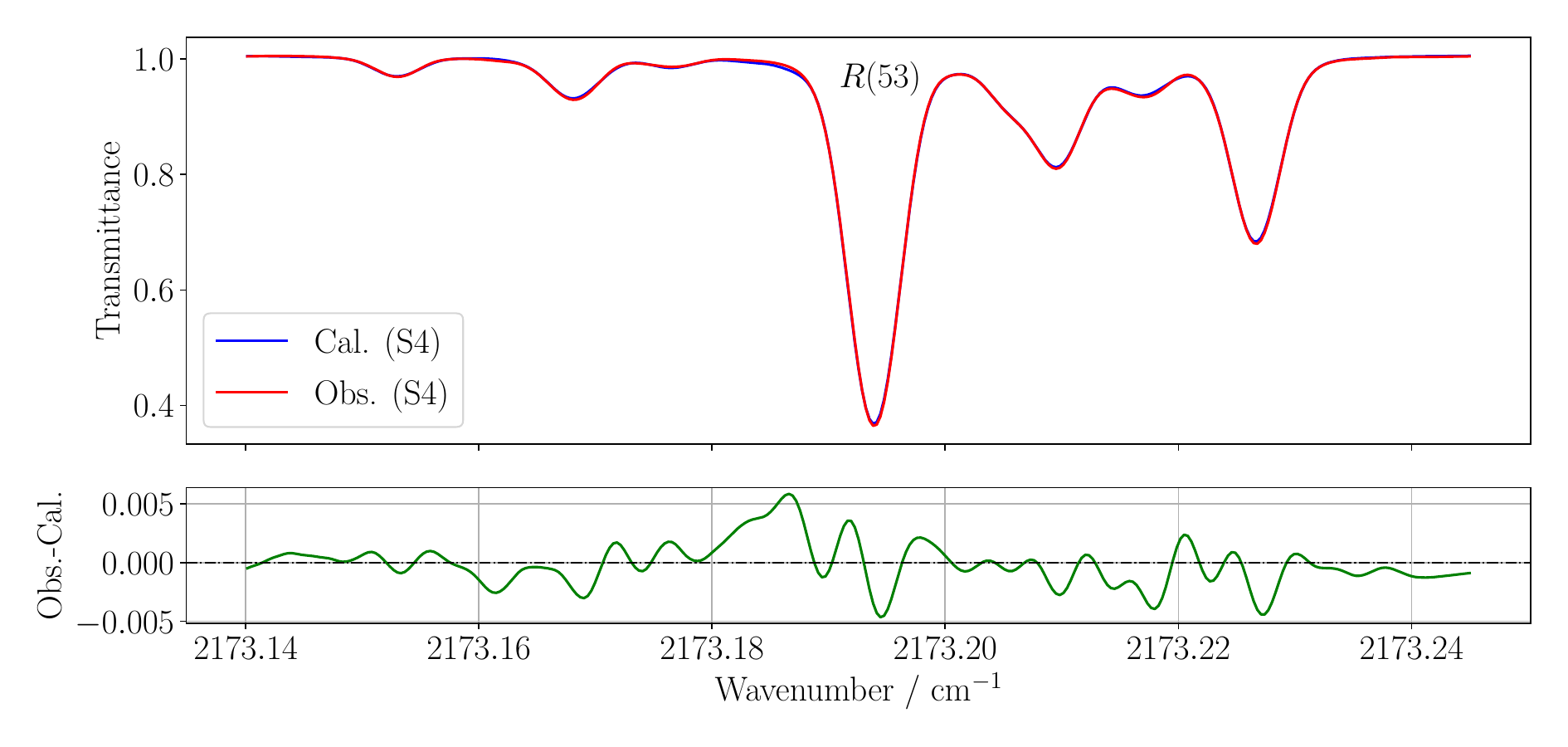}
        \subcaption{}
    \end{minipage}
    \caption{Simulation (in red) and observation (in blue) of the $\nu_3$ $R$(53) line of $\ce{C_2N_2}$ in the S2 (a), S3 (b), S4 (c)} spectra. Residuals are shown in green for each case. 
    \label{fig:P53}
\end{figure*}

We restricted our intensity analysis to the $\nu_3$ band, as the cometary fluorescence model will focus only on this band. The fit was done using the Multi-Spectrum Fitting Program (MSFP), developed by Jean Vander Auwera at Université Libre de Bruxelles. This program takes as input the spectra recorded under different conditions, the total internal partition function of the considered species, and the linelists to be fitted. We measured the intensity (in usual HITRAN units, \SI{}{\per\centi\meter}/(molecule\ $\mathrm{cm^{-2}}$), at $T_{\mathrm{ref}}=296\ K$) and self-broadening coefficients (in $\mathrm{cm^{-1}/atm}$) of 147 lines of the $\nu_3$ band by fitting Voigt profiles. As an example, the line $R$(53) is represented for three different pressures (S2, S3, S4) in Fig. \ref{fig:P53}. We did not include lines known to be blended with absorption lines originating from the three hot bands discussed earlier because the measured line intensities and broadening are overestimated in those cases. From the individual measurements of $S_J$, we derived the total band strength $S_{\nu_3}$ using the intensity formula (equation \ref{eq:int}): $S_{\nu_3}=18.328\pm 0.071~\mathrm{cm^{-2}atm^{-1}}$ ($S_{\nu_3}=18.5~\mathrm{cm^{-2}atm^{-1}}$ in \cite{bmorvan1985}). Contrary to individual line intensities, the band strength is expressed in $\mathrm{cm^{-2}atm^{-1}}$. The conversion between the two systems of units was done using the factors given in \cite{rao2012chapter3}.
The first term in the Herman-Wallis correction was also derived, and we found $A_1=53.4\pm 6.2 \times 10^{-5}$. The comparison between the measured intensities and the calculated intensities is represented in Fig. \ref{fig:int}. The intensity fit has an rms deviation equal to 6.383\%.

\section{Development of a cometary fluorescence model}\label{sec:fluo}
\subsection{Ground-state population and energy levels}
From our analysis, we first computed the energy levels $E_J$ of cyanogen in the ground state. As a parent species, it is assumed that the initial population of $\ce{C_2N_2}$ is mainly in the ground state. This is because the temperature in the coma is very low (typically a few tens of kelvins), and the sublimation of the parent species from the nucleus to the coma does not excite the molecules to higher vibrational states. Moreover, in a first good approximation, collisions are negligible because the coma is an optically thin environment. The initial population distribution in the ground state follows:
\begin{equation}
    N_J=\varepsilon_J\ (2J+1)\ e^{-hcE_J/kT},
\end{equation}
and the initial fractional population $f_J$ is the ratio between $N_J$ and the total partition function $Q$ at temperature $T$.
\begin{equation}
    f_J=\frac{N_J}{Q}.
\end{equation}

\subsection{Branching ratio from the ground state to $\nu_3$}
The population of the upper-state $\nu_3$ is determined using branching ratios, defined as
\begin{align}
    Br(J)&=\frac{L_J}{2J+1} \times (1+A_1m)^2,
\end{align}
which we develop as
\begin{align}
    Br(J;\Delta J =-1)    &= \frac{J}{2J+1}\times (1-A_1J)^2,\label{eq:br1}\\
    Br(J;\Delta J =+1)    &= \frac{J+1}{2J+1}\times (1+A_1(J+1))^2.\label{eq:br2}
\end{align}
In absorption, $\Delta J=-1$ corresponds to the $P$ branch, while $\Delta J=+1$ corresponds to the $R$ branch. The populated $J$ level in the ground state is split to $J$ ($P$ branch) or $J+1$ ($R$ branch) level of the $\nu_3$ excited level, following the usual selection rules for rotovibrational transitions of linear molecules:
\begin{align}
    \Delta \Lambda = &~0,\pm1,\nonumber\\
    \Delta J =&\pm1,&\mathrm{if~\Delta\Lambda=0,~\Lambda=0}, \\
    \Delta J =&~0,\pm1&\mathrm{else}. \nonumber
\end{align}
In our case, $\nu_3$ and GS being both $\Sigma$ states, $\Delta\Lambda=0,~\Lambda=0$, and the $Q$ branch is forbidden. The fractional population $f_{J'}$ in the excited state $J'$ is then
\begin{align}
    f_{J'}=&Br(J=J'-1;\Delta J =+1)f_{J=J'-1},\\
           +&Br(J=J'+1;\Delta J =-1)f_{J=J'+1}.
\end{align}
As an example, the excited $J'=1$ level is populated from the ground states $J=0$ and $J=2$, and
\begin{align}
    f_{J'=1}=&Br(J=0;\Delta J =+1)f_{J=0},\\
    +&Br(J=2;\Delta J =-1)f_{J=2}.
\end{align}
\begin{table}[]
\centering
\resizebox{0.4\linewidth}{!}{%
\begin{tabular}{rrrrr}

\toprule

& \multicolumn{2}{c}{GS} & \multicolumn{2}{c}{$\nu_3$} \\
\cmidrule(lr){2-3} \cmidrule(lr){4-5}
J/J'                      & \multicolumn{1}{c}{$E_J$} & \multicolumn{1}{c}{$f_J$}  & \multicolumn{1}{c}{$E_{J'}$} & \multicolumn{1}{c}{$f_{J'}$} \\

\midrule
0  & 0.00  & 0.00600  & 2157.82 & 0.00297 \\
1  & 0.31  & 0.00892  & 2158.14 & 0.01767 \\
2  & 0.94  & 0.02921  & 2158.76 & 0.01446 \\
3  & 1.89  & 0.01990  & 2159.70 & 0.03943 \\
4  & 3.14  & 0.04936  & 2160.96 & 0.02446 \\
5  & 4.71  & 0.02883  & 2162.52 & 0.05717 \\
6  & 6.60  & 0.06455  & 2164.40 & 0.03202 \\
7  & 8.80  & 0.03496  & 2166.59 & 0.06941 \\
8  & 11.31 & 0.07370  & 2169.10 & 0.03662 \\
9  & 14.14 & 0.03797  & 2171.91 & 0.07552 \\
10 & 17.28 & 0.07668  & 2175.04 & 0.03817 \\
11 & 20.74 & 0.03801  & 2178.49 & 0.07577 \\
12 & 24.51 & 0.07414  & 2182.25 & 0.03699 \\
13 & 28.59 & 0.03560  & 2186.32 & 0.07114 \\
14 & 32.99 & 0.06738  & 2190.70 & 0.03371 \\
15 & 37.70 & 0.03145  & 2195.40 & 0.06302 \\
16 & 42.73 & 0.05793  & 2200.41 & 0.02907 \\
17 & 48.07 & 0.02635  & 2205.73 & 0.05298 \\
18 & 53.72 & 0.04734  & 2211.36 & 0.02384 \\
19 & 59.69 & 0.02101  & 2217.31 & 0.04241 \\
20 & 65.97 & 0.03687  & 2223.57 & 0.01864 \\

\bottomrule
\end{tabular}%
}
\caption{Energy levels and fractional population of $\ce{C_2N_2}$ in the ground state (GS) and in the $\nu_3$ excited state, with $T=50$~K.}
\label{tab:fJ_combined}
\end{table}

Finally, the population that falls back from the excited $J'$ level to the $J''$ level in the ground state through the $R$ branch or $P$ branch is
\begin{align}
    f_{J'',R}=Br(J'=J''+1;\Delta J =-1)f_{J'=J''+1},\\
    f_{J'',P}=Br(J'=J''-1;\Delta J =+1)f_{J'=J''-1},
\end{align}
where $\Delta J$ is now defined as $J''-J'$. This unusual choice in spectroscopy (where $\Delta J$ is normally $J'-J''$) was made to stay consistent with equations \ref{eq:br1} and \ref{eq:br2}. The corresponding emission $g$-factors are obtained by multiplying $f_{J''}$ by the $g$-factor of the band (defined in equation \ref{eq:gnu}):
\begin{align}
    g_{J'',R}=f_{J'',R}~\times~g_{\nu},\\
    g_{J'',P}=f_{J'',P}~\times~g_{\nu}.
\end{align}

    \subsection{Excitation rates}
\begin{figure*}
    \centering
    \includegraphics[width=0.8\linewidth]{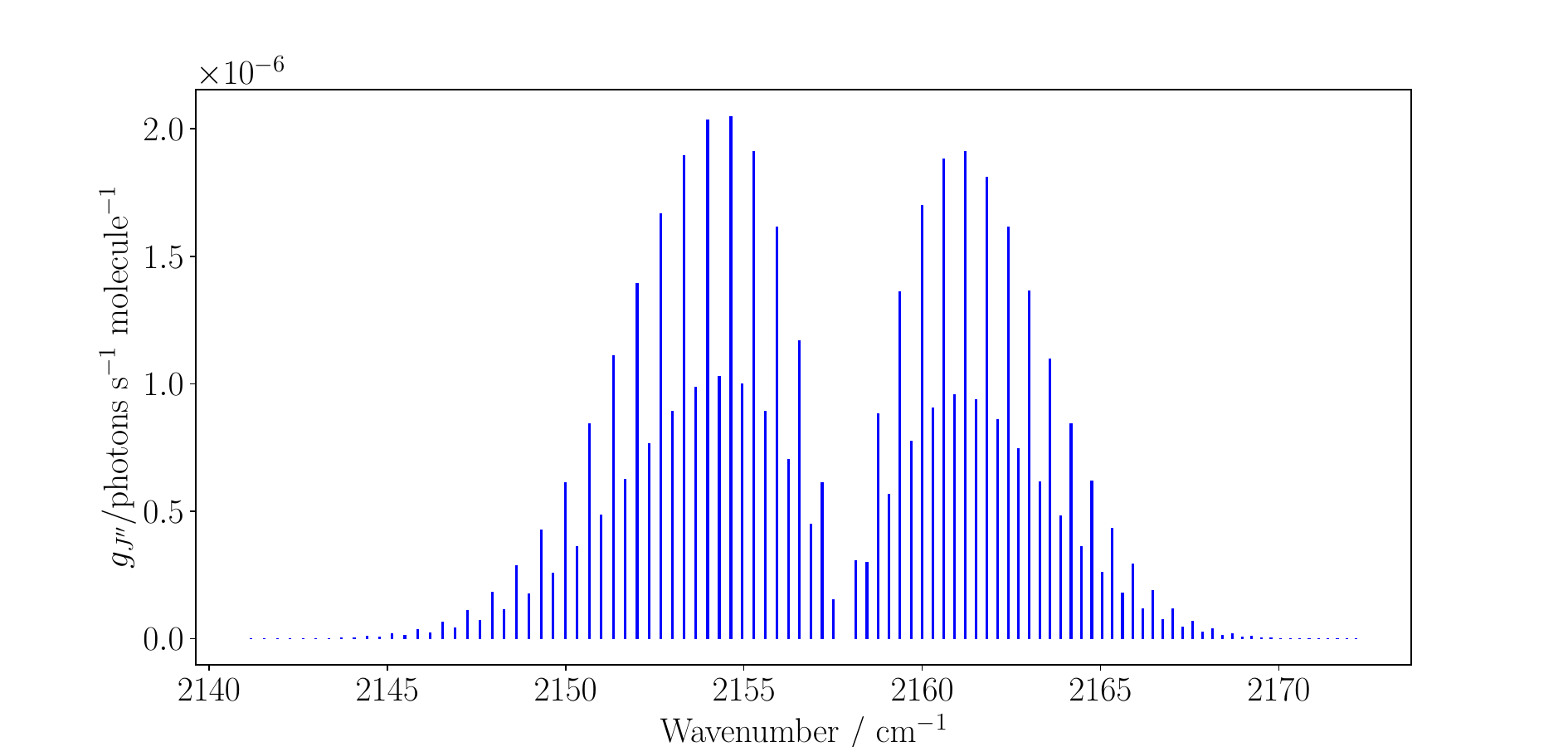}
    \caption{Emission $g$-factors of \ce{C_2N_2} expressed in photons $\mathrm{s^{-1}\ molecule^{-1}}$ at $T$=50 K. The values are the same as in Table \ref{tab:gfactors}.}
    \label{fig:gfactor}
\end{figure*}

     For the $\nu_3$ band, considering an excitation by the Sun at 1 au, $g_{\nu}=5.22\times10^{-5}$ photons $\mathrm{s^{-1}molecule^{-1}}$, a value 50\% higher than the one reported in \cite{bmorvan1985}, despite reporting a similar band strength. We explain this difference by an underestimation of the vibrational partition function $Q_V$ in \cite{bmorvan1985}, impacting the Einstein coefficient of the band and therefore the $g$-factor of the band (respectively, equations \ref{eq:einstein} and \ref{eq:gnu}). At 300 K, $Q_v=1.819$ in \cite{bmorvan1985}, while we have $Q_v=2.703$ and \cite{fayt} have $Q_v=2.70687$. Excitation rates up to $J=20$ for $T=50$ K are presented in Table \ref{tab:gfactors}, and represented in Fig. \ref{fig:gfactor}.
\begin{table}[]
\centering
\resizebox{\linewidth}{!}{%
\begin{tabular}{lrr|lrr}\hline
Line    & Wavenumber/$\mathrm{cm^{-1}}$& $g_{J''}/10^{-6}\mathrm{s^{-1}molecule^{-1}}$  &   Line    & Wavenumber/$\mathrm{cm^{-1}}$& $g_{J''}/10^{-6}\mathrm{s^{-1}molecule^{-1}}$              \\\hline
$P$(20)   & 2151.3387    & 1.111 & $R$(0)    & 2158.1374    & 0.308                         \\
$P$(19)   & 2151.6730    & 0.626 & $R$(1)    & 2158.4494    & 0.303                         \\
$P$(18)   & 2152.0063    & 1.395 & $R$(2)    & 2158.7604    & 0.885                         \\
$P$(17)   & 2152.3386    & 0.768 & $R$(3)    & 2159.0703    & 0.570                         \\
$P$(16)   & 2152.6697    & 1.669 & $R$(4)    & 2159.3791    & 1.364                         \\
$P$(15)   & 2152.9999    & 0.896 & $R$(5)    & 2159.6869    & 0.776                         \\
$P$(14)   & 2153.3289    & 1.897 & $R$(6)    & 2159.9936    & 1.703                         \\
$P$(13)   & 2153.6569    & 0.990 & $R$(7)    & 2160.2992    & 0.907                         \\
$P$(12)   & 2153.9839    & 2.037 & $R$(8)    & 2160.6038    & 1.885                         \\
$P$(11)   & 2154.3097    & 1.031 & $R$(9)    & 2160.9072    & 0.959                         \\
$P$(10)   & 2154.6346    & 2.053 & $R$(10)   & 2161.2097    & 1.914                         \\
$P$(9)    & 2154.9583    & 1.002 & $R$(11)   & 2161.5110    & 0.939                         \\
$P$(8)    & 2155.2810    & 1.916 & $R$(12)   & 2161.8113    & 1.813                         \\
$P$(7)    & 2155.6026    & 0.893 & $R$(13)   & 2162.1105    & 0.862                         \\
$P$(6)    & 2155.9232    & 1.617 & $R$(14)   & 2162.4086    & 1.618                         \\
$P$(5)    & 2156.2427    & 0.705 & $R$(15)   & 2162.7056    & 0.749                         \\
$P$(4)    & 2156.5612    & 1.171 & $R$(16)   & 2163.0016    & 1.368                         \\
$P$(3)    & 2156.8785    & 0.452 & $R$(17)   & 2163.2965    & 0.617                         \\
$P$(2)    & 2157.1948    & 0.614 & $R$(18)   & 2163.5903    & 1.101                         \\
$P$(1)    & 2157.5101    & 0.155 & $R$(19)   & 2163.8830    & 0.485                         \\\hline
\end{tabular}%
}
\caption{Wavenumbers and excitation rates ($g$-factors, noted $g_{J''}$) of 40 lines with $J\leq20$  calculated at $T=50$~K and considering a Sun illumination at 1 au. Excitation rates are expressed in photons $\mathrm{s^{-1}molecule^{-1}}$. The excitation rate of the whole $\nu_3$ band is $g_{\nu}=5.22\times10^{-5}$ photons $\mathrm{s^{-1}molecule^{-1}}$.}
\label{tab:gfactors}
\end{table}

\section{Investigation in comet C/2022 E3 (ZTF)}\label{sec:results}

\subsection{Observations and data reduction}
The Oort cloud comet C/2022 E3 (ZTF) was discovered on 2022 March 2 by the Zwicky Transient Facility. It reached perihelion in 2023 January, at around 1.11 AU, with a peak magnitude of 5 \citep{bolin}. The comet was observed by JWST between 2023 February 28 and 2023 March 4, with $r_h\sim1.3$ au and $\Delta\sim$ 0.9 au \citep{milam}. We investigated the JWST/NIRSPEC spectra obtained with the G395H/F290LP grating/filters configuration, covering the 2.87-5.27 \SI{}{\micro\meter} range through the Integral Field Unit, with a $3.0''\times3.0''$ field of view with the NRSIRS2RAPID readout and a 4-point dither \citep{Boker_2023}. The spectra were extracted with a 1.3" diameter. Third-level processed data were retrieved from the ESA JWST Science Archive\footnote{\url{https://jwst.esac.esa.int/archive/}}.
\subsection{Search for cyanogen}
\begin{figure}
    \centering
    \includegraphics[width=0.6\linewidth]{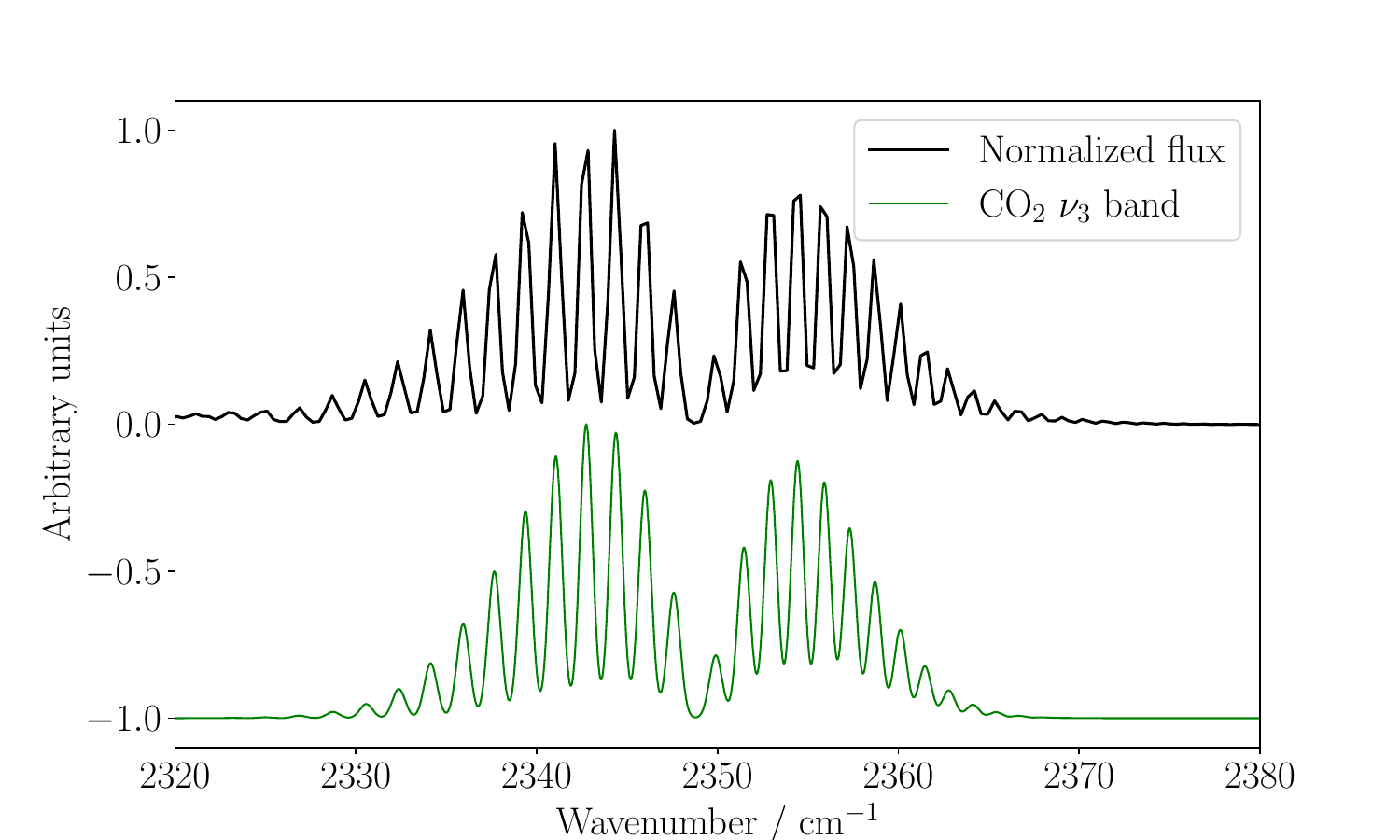}
    \caption{JWST/NIRSPEC spectrum of comet C/2022 E3 (ZTF) (in black) between 2320 and 2380 \SI{}{\per\cm}. A first-order background signal (due to the scattered light on the dust) has been removed, and the spectrum has then been normalized. The intensity distribution in the band enabled us to estimate the rotational temperature of \ce{CO_2} as $T=56$ K. In green, a synthetic fluorescence spectrum at this rotational temperature, based on data from HITRAN \citep{hitran20} and convolved taking into account the resolving power of JWST/NIRSPEC ($R\sim2700$) is represented.}
    \label{fig:CO2nu3}
\end{figure}

The temperature of the coma was estimated by analyzing the intensity distribution of the $\nu_3$ band of \ce{CO_2}, the strongest emission feature in the spectrum. We assumed the gas to be in local thermodynamic equilibrium (LTE), and we expect the relative population levels to follow a Boltzmann law. From HITRAN database \citep{hitran20} and simulation of the fluorescence spectrum in PGOPHER \citep{WESTERN2017221}, we derived $T=56$ K. This is in agreement with the gas temperature derived from radio spectroscopy of the comet in early February by \cite{biver}, where $T=60$ K. The $\nu_3$ band of \ce{CO_2} and the simulation are represented in Fig. \ref{fig:CO2nu3}. We observe that the assumed Boltzmann distribution matches very well the cometary spectrum, confirming the LTE assumption. As $\mathrm{CO_2}$ is a parent molecule with a short lifetime ($\tau=5\times10^5$ s, \cite{crovisier_encrenaz_1983}), it is more concentrated around the nucleus, and no temperature gradient is expected. The lifetime of $\mathrm{C_2N_2}$ being even shorter, the LTE assumption holds for this molecule as well. 

The production rate \( Q \) (in molecules \SI{}{\per\second}) can be expressed as
\begin{equation}
    Q = \frac{4\pi \Delta^2 F}{\tau g h c \nu f}G\label{eq:Q},
\end{equation}
where \( \Delta \) is the geocentric distance (in meters), \( F \) is the flux of the emission line (in \( \mathrm{W\,m^{-2}} \)), \( \tau \) is the lifetime (in seconds) of the molecule, \( g \) is the fluorescence efficiency (in photons per molecule \SI{}{\per\second}) of the emission line, \( h c \nu \) is the energy (in joules) of a photon with a wavenumber $\nu$, and \( f \) is a geometric term representing the fraction of molecules sampled through the telescope aperture relative to the total number of molecules in the coma. We write $G$ the growth factor of the considered species, which is the ratio between the terminal production rate and the nucleus-centered production rate \citep{Russo98}. In Eq. \ref{eq:Q}, $\tau$ and $g$ both depend on the heliocentric distance ($R_h$, expressed in au). However, as $g \propto R_h^{-2}$ (see Eq. \ref{eq:gnu}) and $\tau \propto R_h^{2}$, their product is independent of the heliocentric distance.

To quantify an upper limit for $Q$(\ce{C_2N_2}) in comet C/2022 E3 (ZTF), we used the Haser model \citep{haser}, which describes the distribution of chemical species in the coma. It relies on the following hypothesis:
\begin{enumerate}
    \item The cometary nucleus is assumed to be spherical, with a radius $R_0$.
    \item Parent species are isotropically ejected away from the nucleus, with a constant velocity noted $v_0$.
    \item The photodissociation of parent species is given by the law
    \begin{equation}
        n=n_0\ e^{-t/\tau},
    \end{equation}
    where $n_0$ is the number of molecules at time $t=0$, and $\tau$ is the photodissociation lifetime of the corresponding molecule.
\end{enumerate}

The density of parent species at a distance $R>R_0$ from the nucleus is given by
\begin{equation}
    n(R)=\frac{Q}{4\pi v_0R^2}\exp\left(-\frac{R-R_0}{\gamma}\right),
\end{equation}
where $Q$ is the production rate of the parent species (expressed in molecules~\SI{}{\per\second}) and $\gamma$ is the scale length, defined as $\gamma=v_0\tau_0$. The column density of molecules, $N(\rho)$, is obtained by integrating $n(R)$ along the line of sight:
\begin{align}
    N(\rho)&=\frac{Q}{4\pi v_0}\exp\left(\frac{R_0}{\gamma}\right)\int_{-\infty}^{\infty}\frac{1}{R^2}\exp\left(-\frac{R}{\gamma}\right)dz\\ \label{eq:integral},
    &=\frac{Q}{2\pi v_0 \rho}\exp\left(\frac{R_0}{\gamma}\right)\int_0^{\pi/2}\exp\left(-\frac{\rho}{\gamma \cos{\alpha}}\right)d\alpha,
\end{align}
where we changed the variable of integration for $\alpha$ such that $\rho=R\cos{\alpha}$, and $dz=\rho d\alpha/\cos^2{\alpha}$ (the change of variables is illustrated in Fig. \ref{fig:chgtvariable}). We write $M(s)$ for the total number of molecules for a specific parent species observed through a circular aperture $s$ as 
\begin{equation}
    M(s)=2\pi\int_{R_0}^R\rho N(\rho)d\rho\label{eq:M(s)},
\end{equation}
where we assumed that the aperture was centered on the nucleus. $R\approx\Delta\times s$ is the physical size at the comet given the geocentric distance $\Delta$, with $s$ in radian.
The total number of molecules is obtained when $s,R\to \infty$ in Eq. \ref{eq:M(s)}:
\begin{align}
    M(\infty) &= \frac{Q}{v_0} \exp{\left(\frac{R_0}{\gamma}\right)} \left(\gamma - \int_0^{R_0}  \int_0^{\pi/2} \exp{\left(-\frac{\rho}{\gamma \cos{\alpha}}\right)}  d\alpha d\rho\right),\\
    &\approx \frac{Q\gamma}{v_0}, \quad \text{since } R_0 \ll \gamma.
\end{align}
For a rectangular aperture of dimensions $s_1$, $s_2$, centered on the nucleus,
\begin{equation}
    M(s_1,s_2)=\left(4\int_0^{R_2}\int_0^{R_1}N(x,y)dxdy\right)-\left(2\pi\int_0^{R_0}\rho N(\rho)d\rho\right),
\end{equation}
where $R_1\approx\Delta\times s_1$, $R_2\approx\Delta\times s_2$. Again, as the second term is much smaller than the first one, the total number of molecules is approximately
\begin{equation}
    M(s_1,s_2)\approx4\int_0^{R_2}\int_0^{R_1}N(x,y)dxdy.
\end{equation}
The geometric factor $f$ in the production rate's equation (Eq. \ref{eq:Q}) is then
\begin{equation}
    f=\frac{M(s_1,s_2)}{M(\infty)}.
\end{equation}
This factor depends on the slit's dimensions ($3.0''\times3.0''$), the radius $R_0$ of the cometary nucleus (we assumed $R_0=2.5$ km, a value compatible with a recent photometric study of C/2022 E3 (ZTF), \cite{liu}), the geocentric distance $\Delta$, the lifetime $\tau$ at the heliocentric distance $R_h$ ($\tau=\tau_{\text{1au}}\times R_h^{2}$) and the ejection velocity $v_0$ of the parent species (we chose $v_0=0.8\times R_h^{-0.5}$ \SI{}{\kilo\meter\per\second}).

The photodissociation rate of \ce{C_2N_2} at 1 au ($k=3.08\times10^{-5}$~\SI{}{\per\second}) was computed by \cite{bmorvan1985}, using vacuum ultraviolet spectra from \cite{nuth}. This leads to a photodissociation lifetime at 1~au equal to $\tau_{\text{1au}}=3.25\times10^4$~s, which is very close to the lifetime of acetylene (\ce{C_2H_2}) at 1 au ($\tau_{\text{1au}}=3.2\times10^4$~s), as reported by \cite{crovisier_encrenaz_1983}

\begin{figure*}
    \centering
    \includegraphics[width=0.8\linewidth]{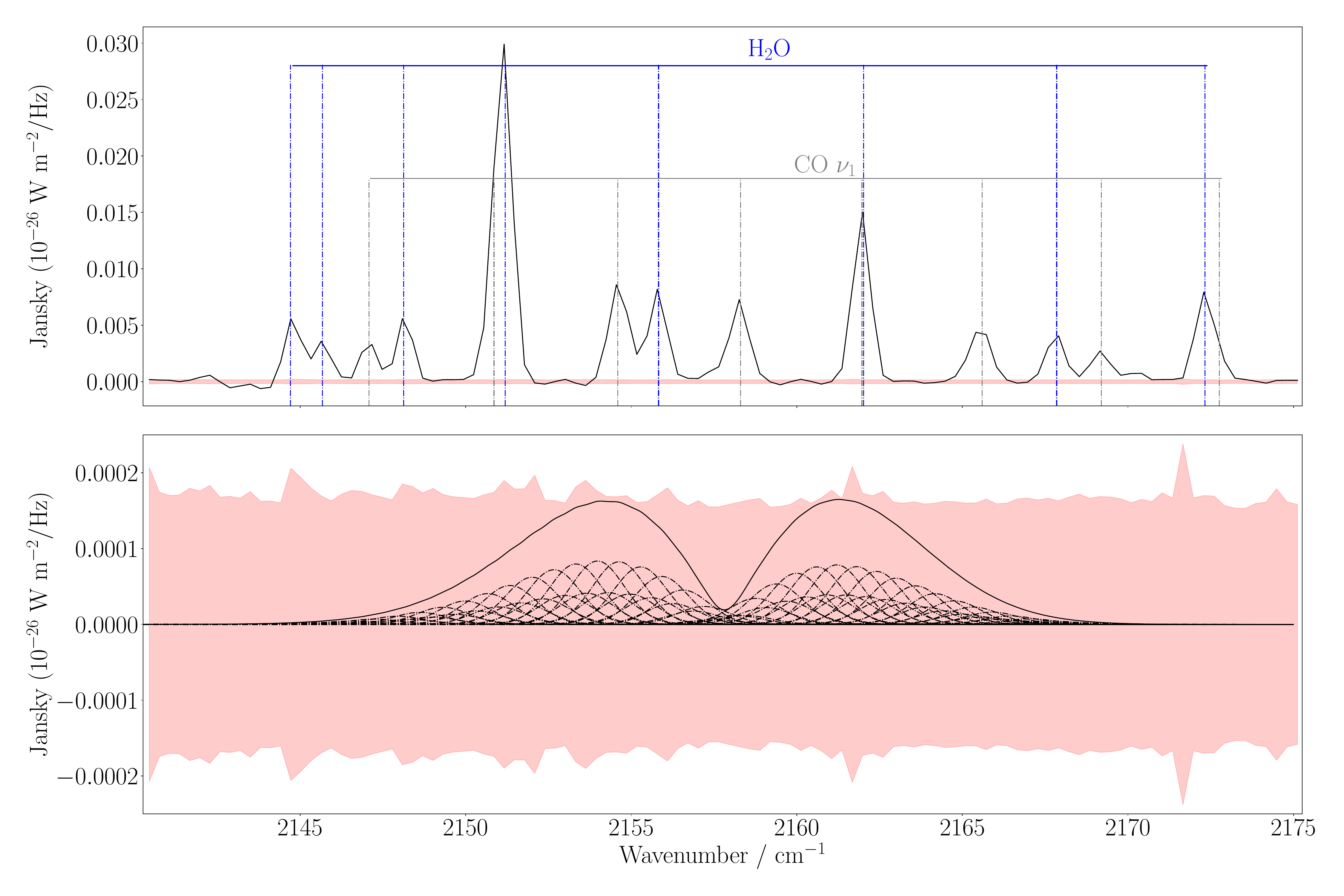}
    \caption{\textit{Top:} JWST/NIRSPEC spectrum of comet C/2022 E3 (ZTF) between 2140 and 2175 \SI{}{\per\centi\meter}. Emission lines of \ce{CO} ($\mathrm{\nu_1}$) and of \ce{H_2O} are represented by alternated gray and blue lines, respectively. \textit{Bottom:} Synthetic spectrum of \ce{C_2N_2} (solid curve) calculated as the sum of $\nu_3$ lines (dashed lines) up to $J=50$ at $T=56$ K, convoluted by the JWST/NIRSPEC resolving power ($R\sim2700$). The $\pm3\sigma$ flux error envelope is represented in pink in each plot.}
    \label{fig:jwst}
\end{figure*}

We simulated synthetic spectra of \ce{C_2N_2} up to $J=50$ to retrieve an upper limit (at 3$\sigma$-level) in the spectrum of C/2022 E3 (ZTF). We assumed the rotational temperature of \ce{C_2N_2} was equal to the rotational temperature derived with the \ce{CO_2} $\nu_3$ band ($T_{rot}=56$ K). We convolved the emission lines using the resolving power of JWST/NIRSPEC, choosing a Gaussian profile for the convolution.
The resolution of JWST/NIRSPEC in this configuration ($R\sim2700$) is too low to observe lines of \ce{C_2N_2} individually, as its rotational constant is low ($B\sim0.157$~\SI{}{\per\centi\meter}). As a comparison, the much larger rotational constant of \ce{CO} ($B\sim1.922$~\SI{}{\per\centi\meter}, \cite{RANK1965418}) leads to a much larger separation between two consecutive lines. This is illustrated in Fig. \ref{fig:jwst}. We obtained $Q(\ce{C_2N_2})<6\times10^{25}$ molec. $\mathrm{s^{-1}}$ ($3\sigma$), where this upper limit corresponds to the maximum production rate at which the total contribution of the \ce{C_2N_2} band remains below the $3\sigma$ noise level in the JWST/NIRSPEC spectrum. Since the $3\sigma$ noise level is calculated from regions
where no emission lines are visible, it takes into account the various noise sources (coming from both the instrument itself and 
photon noise due to the intensity of the light scattered by the dust particles). A blend with other molecular species is unlikely – 
with the exception of the bright emission lines appearing in Figure \ref{fig:jwst} – because no emission bands appear in the 
spectral range corresponding to the \ce{C_2N_2}  band.

From three water lines with known g-factors at $T$=60~K (summarized in Table \ref{tab:leau}) \citep{DELLORUSSO2000324}, we estimated the production rate of water of C/2022 E3 (ZTF) on 01/03/23 as $Q(\ce{H_2O})\sim3\times10^{28}$ molec. $\mathrm{s^{-1}}$. We assumed a usual ortho to para ratio of 3, and a growth factor of 1.8. The derived value is very close to the mean production rate of water derived by \cite{combi} (reported in \cite{Li_2025}) one month prior to JWST measurements, when $R_h=1.162$ au (post-perihelion): $3.05\pm0.06 \times10^{28}$ molec. $\mathrm{s^{-1}}$. 

\begin{table*}[]
\resizebox{\linewidth}{!}{%
\begin{tabular}{llllll} \hline
Band assignment & Line assignment & Rest frequency ($\mathrm{cm^{-1}}$) & g-Factors at 60K ($\mathrm{s^{-1}}$) & Measured flux ($\mathrm{Wm^{-2}}$) & $Q_{H_2O}$ ($\mathrm{s^{-1}}$)\\ \hline
100-010         & $1_{01}-1_{10}$ & 2039.95    & $7.15\times 10^{-7}$     & $2.9631(52)\times 10^{-18}$  & $2.96\times10^{28}$\\
001-010         & $0_{00}-1_{01}$ & 2137.37    & $8.22\times 10^{-7}$     & $3.6173(46)\times 10^{-18}$  & $3.00\times10^{28}$\\
001-010         & $1_{11}-1_{10}$ & 2151.19    & $5.70\times 10^{-7}$     & $2.5353(48)\times 10^{-18}$  & $3.01\times10^{28}$\\  \hline                                 
\end{tabular}
}
\caption{Emission lines of water used to estimate the water production rate in comet C/2022 E3 (ZTF) on 01/03/23. g-Factors (from \cite{DELLORUSSO2000324}) are expressed for $R_h=1\ \mathrm{au}$, and assuming an ortho to para ratio of 3.}
\label{tab:leau}
\end{table*}

Our upper limit on the mixing ratio of cyanogen is \ce{C_2N_2}/\ce{H_2O} $<0.22$\% (3$\sigma$). This value is still higher than the abundances derived \textit{in situ} in the coma of 67P \citep{Hanni_Altwegg_Balsiger_Combi_Fuselier_De_Keyser_Pestoni_Rubin_Wampfler_2021}: \ce{C_2N_2}/HCN = $0.18 \pm 0.09\%$, (1.24-1.74 au), and HCN/\ce{H_2O} = $0.14 \pm 0.04\%$ in May 2015, at the same time of detection of \ce{C_2N_2} \citep{rubin2019}.

\subsection{Perspectives}

The mixing ratios measured in comet 67P mentioned above correspond to a \ce{C_2N_2}/\ce{H_2O}$\sim 3\times 10^{-6}$ for this comet. Such a ratio, if similar for other comets would imply a minimum water production rate of about $6\times10^{25}/3\times10^{-6}\sim 2\times 10^{31}$ molec. $\mathrm{s^{-1}}$ as the minimum water production rate to allow detection of \ce{C_2N_2} (if considering similar heliocentric and geocentric distances than the ones at the time of comet C/2022 E3 (ZTF) observations). Such a water production rate can be compared to that measured in comet C/1995 O1 Hale-Bopp of about $8.3\times10^{30}$ molec. $\mathrm{s^{-1}}$ \citep{DELLORUSSO2000324}, i.e., of the same order of magnitude. Such values confirm that \ce{C_2N_2} could reasonably be detected by infrared spectroscopy with ground-based facilities or infrared space observatories for some comets, if they are bright enough, for small heliocentric and geocentric distances (those at the time of observations of comet C/2022 E3 (ZTF) not being truly optimal). Additionally, advancements in telescope technology will contribute to reducing uncertainties. Instruments like METIS (Mid-Infrared ELT Imager and Spectrograph), planned for installation in 2028 on the 39-meter ESO Extremely Large Telescope (ELT) \citep{metis}, will offer high-resolution spectroscopy ($R\sim100000$) in the L/M infrared windows.
\cite{brandl2010metis} provide a comparison of the sensitivity between 
JWST/NIRSPEC and METIS for the M band (their Fig. 3). It is found that METIS 
should be about twice as sensitive as JWST/NIRSPEC in this band, which 
should facilitate the detection of \ce{C_2N_2} emission lines, mainly due to 
a much higher spectral resolution allowing the separation of the different lines.
Such an instrument should make it possible to better understand the production rates of \ce{C_2N_2}, thus offering new possibilities for probing its presence in cometary atmospheres.

\section{Conclusions}

This study presents a fluorescence model for the $\nu_3$ vibrational band of cyanogen (\ce{C_2N_2}), offering a robust tool for analyzing its spectral characteristics in cometary environments. By combining high-resolution laboratory spectra with detailed theoretical modeling, we have, for the first time, derived line-by-line excitation rates for this molecule. The analysis of the $\nu_3$ band confirms that cyanogen's fluorescence efficiency is low; however, future bright comets observed either with JWST or ground-based facilities such as METIS are expected to significantly improve the detection and characterization of cyanogen in cometary spectra.

Cyanogen is the second molecule studied as part of the COSMIC project (COmputation and Spectroscopy of Molecules in the Infrared for Comets), following chloromethane (\ce{CH_3Cl}, \cite{HARDY2023108779}). Although the ESA/Rosetta mission led to the discovery of approximately 40 new molecules, molecular databases remain incomplete, lacking data on many key species needed to analyze trace elements in cometary spectra effectively. This work highlights the importance of continued laboratory spectroscopy efforts to expand these databases and enhance our understanding of cometary chemistry, and of the origins of our solar system.

\section{Data availability}
The database supporting the results presented in this article is freely available at the VAMDC \url{https://vamdc.icb.cnrs.fr/download/C2N2-Dijon-Database.zip}. A copy of this databse has also been deposited to Zenodo: \url{https://doi.org/10.5281/zenodo.15658823}. It contains the rotational, vibrational, and total partition function, the analysis of line positions for all bands discussed in this work (summarized in a PGOPHER output file), the intensity fitting of the $\nu_3$ lines performed using the MSFP program, as well as the calculated $g$-factors at various temperatures.

\section{Acknowledgements}
The authors are grateful to Jean Vander Auwera from Université Libre de Bruxelles for providing the MSFIT program to fit lines intensities. The COSMIC project is
financed by the EIPHI Graduate School and held by ICB and UTINAM, supported by the Conseil Régional de Bourgogne Franche-Comté and the French National Research Agency (ANR).

\appendix

\section{Additional figures}
\clearpage
\begin{figure*}
    \centering
    \includegraphics[width=0.5\linewidth]{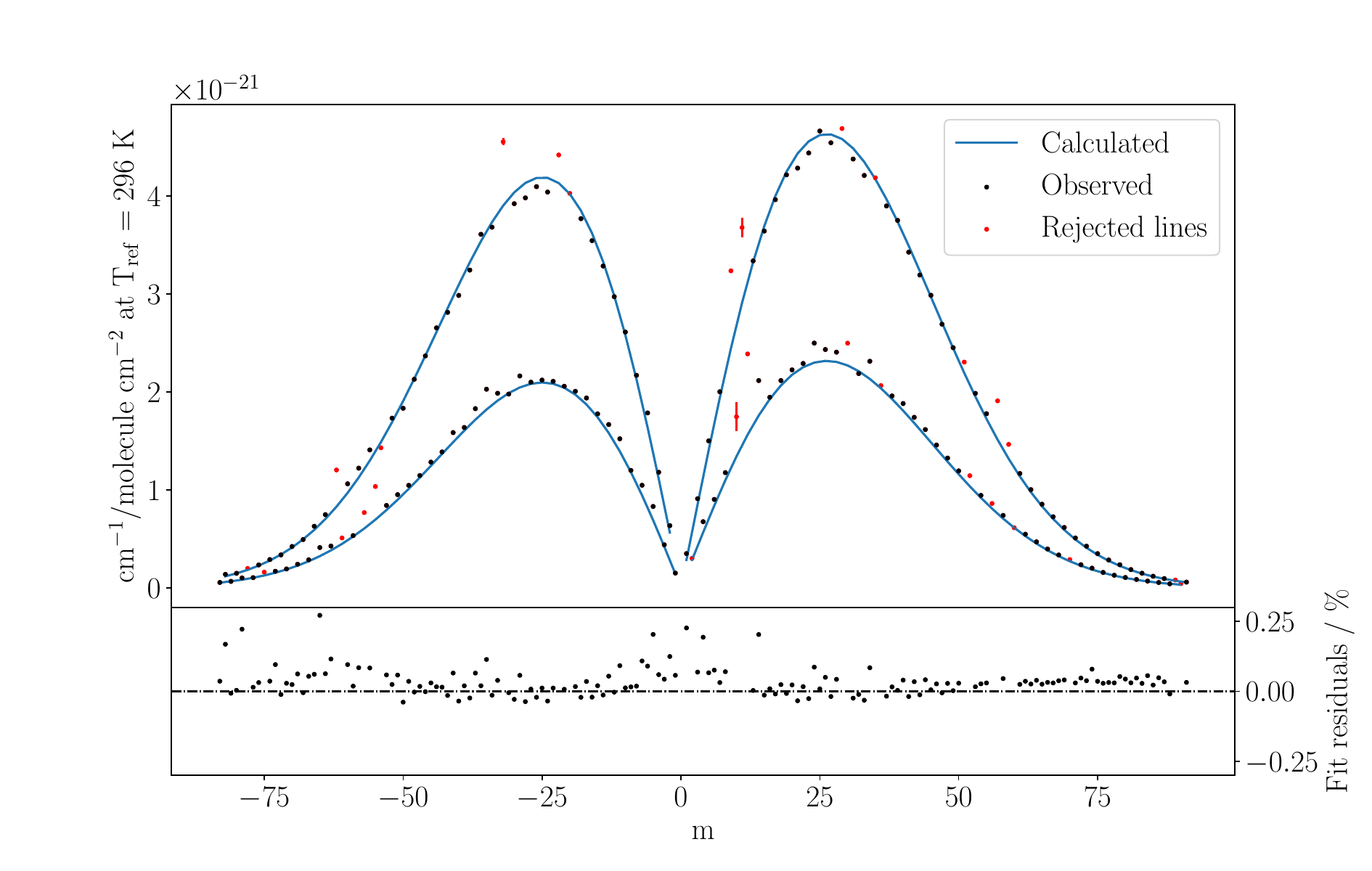}
    \caption{\textit{up} Comparison between measured (dots) and calculated (blue solid lines) intensities of $\nu_3$ absorption lines with respect to m. Black points represent the 147 lines that were used to derive the band strength and the Herman-Wallis correction. We also represented the rejected lines in red: they correspond to lines blended with at least one line originating from a hot band. Error bars are displayed for each point, but are too small to be seen in most of the cases. \textit{bottom} Fit residuals, expressed as the relative difference ((Obs.-Cal.)/Cal.), showing deviations between observed and calculated intensities.}
    \label{fig:int}
\end{figure*}

\begin{figure*}[ht!]
    \centering
    \begin{minipage}{\textwidth}
        \centering
        \includegraphics[width=0.5\textwidth]{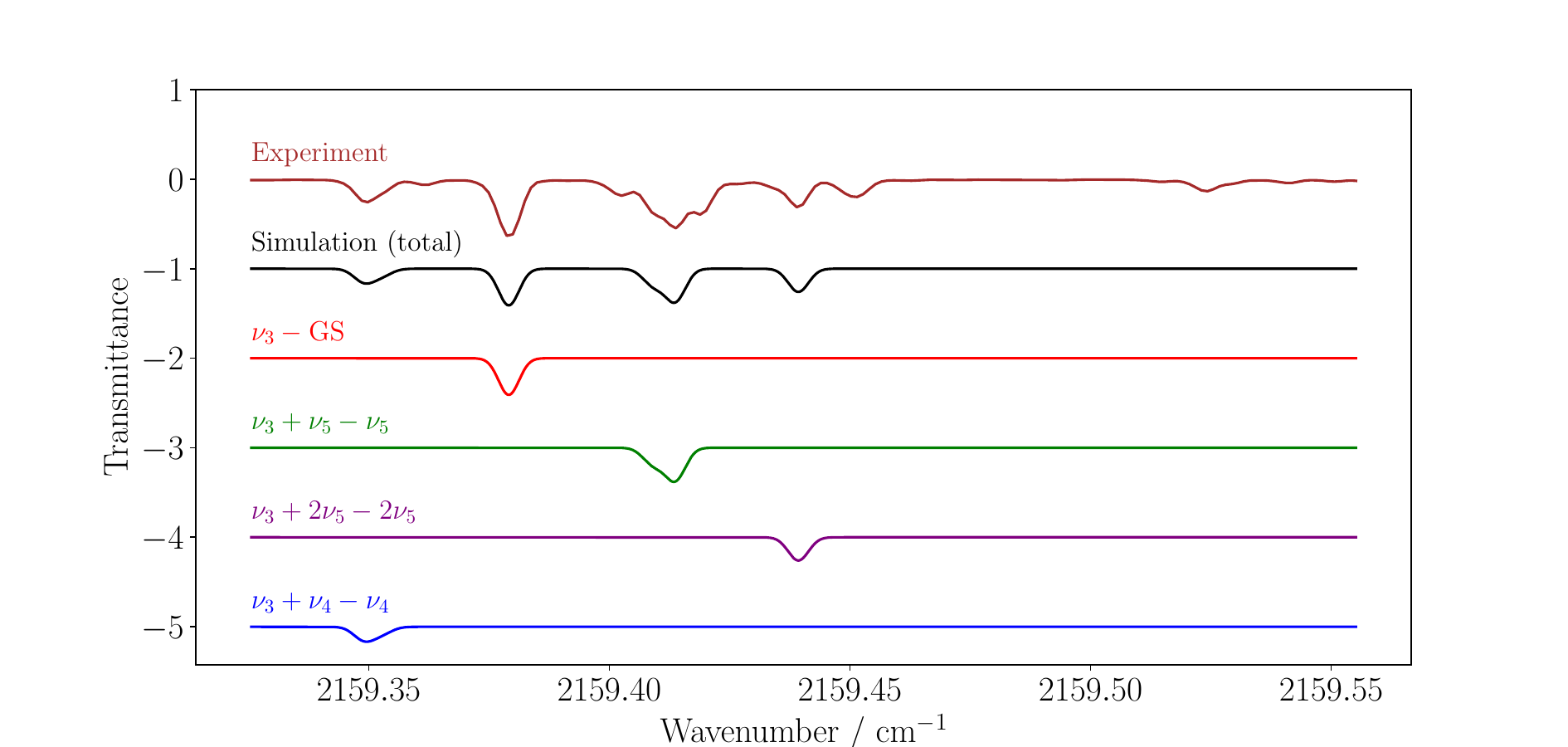}
        \subcaption{Comparison between the calculated and experimental spectrum in the region 2159.33-2159.56 \SI{}{\per\centi\meter}. The $\nu_3+\nu_5-\nu_5$ $R$(10) lines (in green) and the $\nu_3+\nu_4-\nu_4$ $R$(14) lines (in blue) are blended.}
    \end{minipage}
    
    \vspace{0em}
    
    \begin{minipage}{\textwidth}
        \centering
        \includegraphics[width=0.5\textwidth]{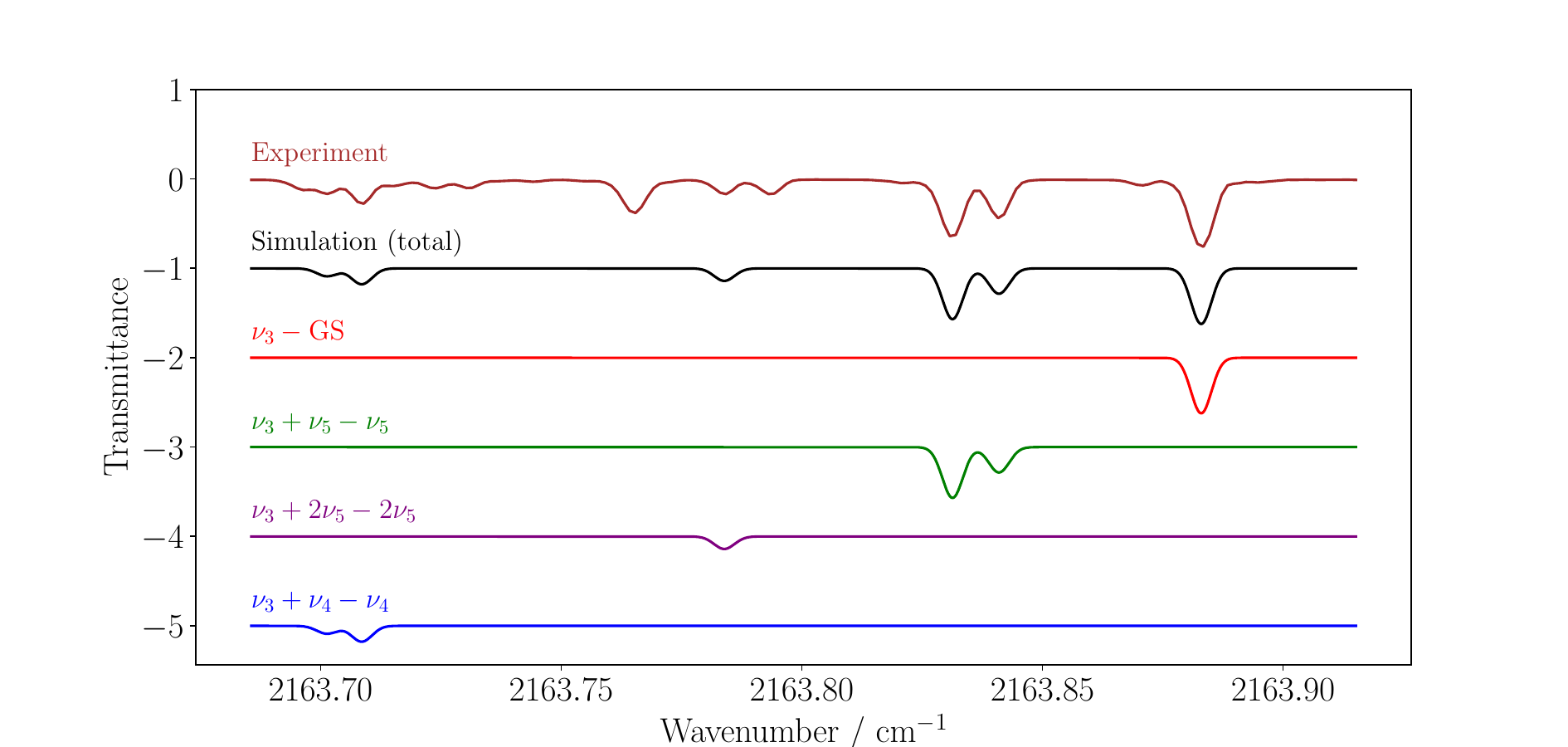}
        \subcaption{Comparison between the calculated and experimental spectrum in the region 2163.69-2163.92 \SI{}{\per\centi\meter}. The $\nu_3+\nu_5-\nu_5$ $R$(25) lines (in green) and the $\nu_3+\nu_4-\nu_4$ $R$(29) lines (in blue) are separating.}
    \end{minipage}
    
    \vspace{0em}
    
    \begin{minipage}{\textwidth}
        \centering
        \includegraphics[width=0.5\textwidth]{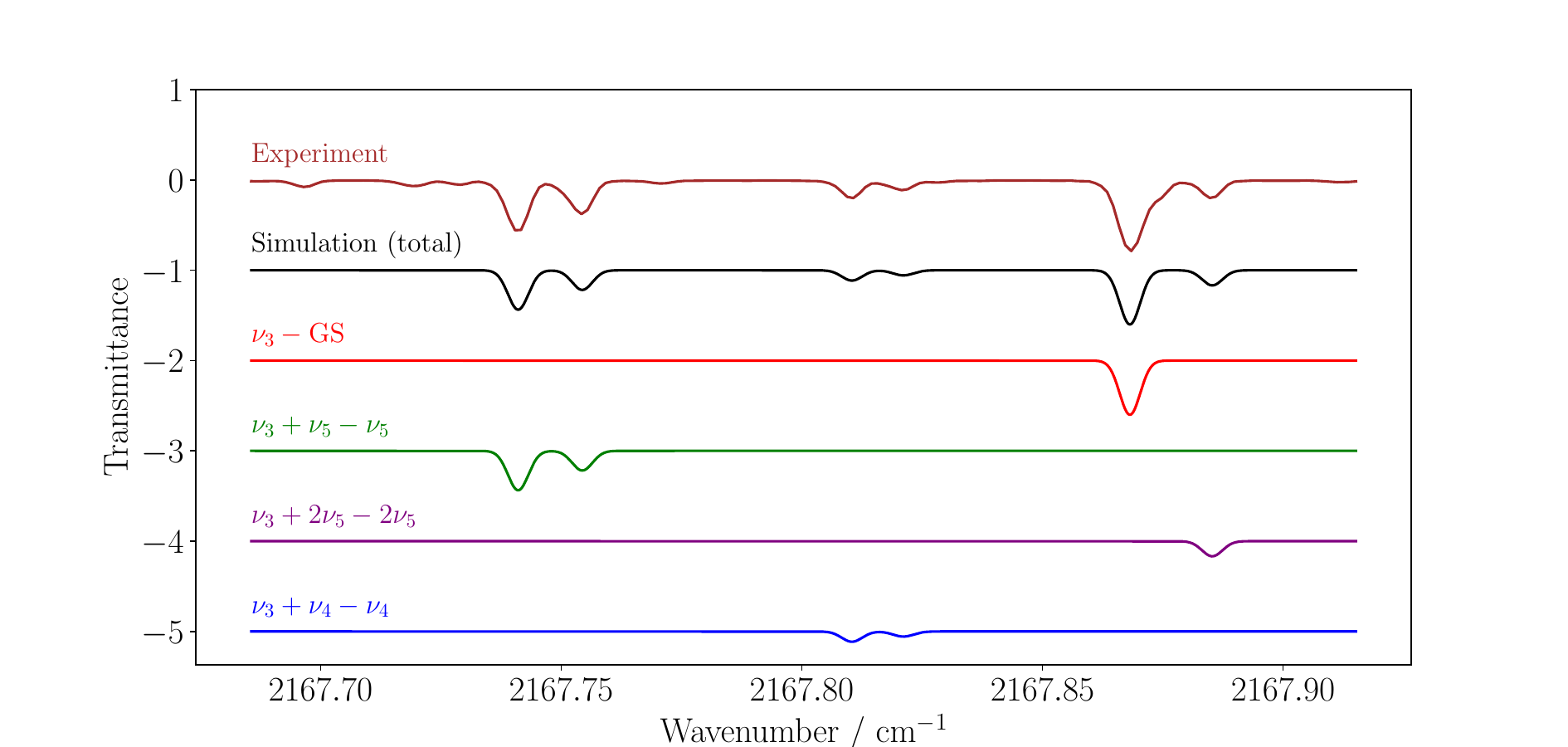}
        \subcaption{Comparison between the calculated and experimental spectrum in the region 2167.69-2167.92 \SI{}{\per\centi\meter}. The $\nu_3+\nu_5-\nu_5$ $R$(39) lines (in green) and the $\nu_3+\nu_4-\nu_4$ $R$(44) lines (in blue) are well separated.}
    \end{minipage}
    
    \caption{Three regions in the $R$-branch with increasing rotational quantum number $J$ (in respect to each band) from (a) to (c). The splitting of $\nu_3+\nu_5-\nu_5$ and $\nu_3+\nu_4-\nu_4$ is getting stronger as $J$ is increasing. Some absorption lines coming from hot bands not taken into account in this work can be observed while comparing the experimental spectrum (in brown) with the total simulation (in black). An arbitrary Gaussian profile and relative intensities have been chosen to calculate the spectra in PGOPHER. }
    \label{fig:lambda}
\end{figure*}

\begin{figure*}
    \centering
    \includegraphics[width=0.5\linewidth]{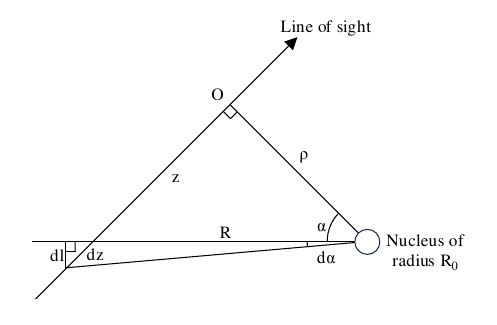}
    \caption{Variables used to derive the integral in equation \ref{eq:integral}. $\rho=R\cos\alpha$, $dl/R=\tan d\alpha\approx d\alpha$, and $dl/dz=\cos\alpha$. This leads to $dz=\rho/\cos^2{\alpha}\ d\alpha$}
    \label{fig:chgtvariable}
\end{figure*}
\clearpage
\bibliographystyle{aasjournal}
\bibliography{biblio}

\begin{thebibliography}{}
\expandafter\ifx\csname natexlab\endcsname\relax\def\natexlab#1{#1}\fi
\providecommand{\url}[1]{\href{#1}{#1}}
\providecommand{\dodoi}[1]{doi:~\href{http://doi.org/#1}{\nolinkurl{#1}}}
\providecommand{\doeprint}[1]{\href{http://ascl.net/#1}{\nolinkurl{http://ascl.net/#1}}}
\providecommand{\doarXiv}[1]{\href{https://arxiv.org/abs/#1}{\nolinkurl{https://arxiv.org/abs/#1}}}

\bibitem[{Altwegg {et~al.}(2019)Altwegg, Balsiger, \& Fuselier}]{Altwegg_Balsiger_Fuselier_2019}
Altwegg, K., Balsiger, H., \& Fuselier, S.~A. 2019, Annual Review of Astronomy and Astrophysics, 57, 113–155, \dodoi{10.1146/annurev-astro-091918-104409}

\bibitem[{{Biver, N.} {et~al.}(2024){Biver, N.}, {Bockelée-Morvan, D.}, {Handzlik, B.}, {Sandqvist, Aa.}, {Boissier, J.}, {Drozdovskaya, M. N.}, {Moreno, R.}, {Crovisier, J.}, {Lis, D. C.}, {Cordiner, M.}, {Milam, S.}, {Roth, N. X.}, {Bonev, B. P.}, {Dello Russo, N.}, {Vervack, R.}, {Opitom, C.}, \& {Kawakita, H.}}]{biver}
{Biver, N.}, {Bockelée-Morvan, D.}, {Handzlik, B.}, {et~al.} 2024, Astronomy and Astrophysics, 690, A271, \dodoi{10.1051/0004-6361/202450921}

\bibitem[{{Bockelee-Morvan} \& {Crovisier}(1985)}]{bmorvan1985}
{Bockelee-Morvan}, D., \& {Crovisier}, J. 1985, Astronomy and Astrophysics, 151, 90

\bibitem[{{Bolin} {et~al.}(2022){Bolin}, {Masci}, {Ip}, {Helou}, {Kramer}, {Lin}, {Prince}, {Sato}, {Paul}, {Yoshimoto}, {Urbanik}, {Denneau}, {Siverd}, {Tonry}, {Weiland}, {Erasmus}, {Fitzsimmons}, {Lawrence}, {Robinson}, {Siverd}, {Tonry}, {Birtwhistle}, {Jacques}, {Hug}, {Korlevic}, {Buzzi}, {Bacci}, {van Buitenen}, {Buczynski}, {Hale}, {Masek}, {Guido}, {Rocchetto}, {Bryssinck}, {Milani}, {Savini}, {Valvasori}, {Ligustri}, {Bacci}, {Maestripieri}, {Tesi}, {Fagioli}, \& {Lutkenhoner}}]{bolin}
{Bolin}, B.~T., {Masci}, F.~J., {Ip}, W.~H., {et~al.} 2022, Minor Planet Electronic Circulars, 2022-F13

\bibitem[{Brandl {et~al.}(2010)Brandl, Blommaert, Glasse, Lenzen, \& Pantin}]{brandl2010metis}
Brandl, B., Blommaert, J., Glasse, A., Lenzen, R., \& Pantin, E. 2010, The Messenger, 140, 30

\bibitem[{{Brandl} {et~al.}(2018){Brandl}, {Absil}, {Ag{\'o}cs}, {Baccichet}, {Bertram}, {Bettonvil}, {van Boekel}, {Burtscher}, {van Dishoeck}, {Feldt}, {Garcia}, {Glasse}, {Glauser}, {G{\"u}del}, {Haupt}, {Kenworthy}, {Labadie}, {Laun}, {Lesman}, {Pantin}, {Quanz}, {Snellen}, {Siebenmorgen}, \& {van Winckel}}]{metis}
{Brandl}, B.~R., {Absil}, O., {Ag{\'o}cs}, T., {et~al.} 2018, in Society of Photo-Optical Instrumentation Engineers (SPIE) Conference Series, Vol. 10702, Ground-based and Airborne Instrumentation for Astronomy VII, ed. C.~J. {Evans}, L.~{Simard}, \& H.~{Takami}, 107021U, \dodoi{10.1117/12.2311492}

\bibitem[{Böker {et~al.}(2023)Böker, Beck, Birkmann, Giardino, Keyes, Kumari, Muzerolle, Rawle, Zeidler, Abul-Huda, de~Oliveira, Arribas, Bechtold, Bhatawdekar, Bonaventura, Bunker, Cameron, Carniani, Charlot, Curti, Espinoza, Ferruit, Franx, Jakobsen, Karakla, López-Caniego, Lützgendorf, Maiolino, Manjavacas, Marston, Moseley, Ogle, Perna, Peña-Guerrero, Pirzkal, Plesha, Proffitt, Rauscher, Rix, del Pino, Rustamkulov, Sabbi, Sing, Sirianni, Plate, Úbeda, Wahlgren, Wislowski, Wu, \& Willott}]{Boker_2023}
Böker, T., Beck, T.~L., Birkmann, S.~M., {et~al.} 2023, Publications of the Astronomical Society of the Pacific, 135, 038001, \dodoi{10.1088/1538-3873/acb846}

\bibitem[{{Combi} {et~al.}(2023){Combi}, {M{\"a}kinen}, {Bertaux}, {Quemerais}, \& {Ferron}}]{combi}
{Combi}, M., {M{\"a}kinen}, T., {Bertaux}, J.-L., {Quemerais}, E., \& {Ferron}, S. 2023, in AAS/Division for Planetary Sciences Meeting Abstracts, Vol.~55, AAS/Division for Planetary Sciences Meeting Abstracts \#55, 322.04

\bibitem[{{Crovisier}(1984)}]{crovisier_1984}
{Crovisier}, J. 1984, Astronomy and Astrophysics, 130, 361

\bibitem[{{Crovisier} \& {Encrenaz}(1983)}]{crovisier_encrenaz_1983}
{Crovisier}, J., \& {Encrenaz}, T. 1983, Astronomy and Astrophysics, 126, 170

\bibitem[{{Dello Russo} {et~al.}(1998){Dello Russo}, {DiSanti}, {Mumma}, {Magee-Sauer}, \& {Rettig}}]{Russo98}
{Dello Russo}, N., {DiSanti}, M.~A., {Mumma}, M.~J., {Magee-Sauer}, K., \& {Rettig}, T.~W. 1998, Icarus, 135, 377, \dodoi{10.1006/icar.1998.5990}

\bibitem[{{Dello Russo} {et~al.}(2000){Dello Russo}, Mumma, DiSanti, Magee-Sauer, Novak, \& Rettig}]{DELLORUSSO2000324}
{Dello Russo}, N., Mumma, M.~J., DiSanti, M.~A., {et~al.} 2000, Icarus, 143, 324, \dodoi{https://doi.org/10.1006/icar.1999.6268}

\bibitem[{Fayt {et~al.}(2012)Fayt, Jolly, Benilan, Manceron, Kwabia-Tchana, \& Guillemin}]{fayt}
Fayt, A., Jolly, A., Benilan, Y., {et~al.} 2012, Journal of Quantitative Spectroscopy and Radiative Transfer, 113, 1195, \dodoi{https://doi.org/10.1016/j.jqsrt.2012.02.003}

\bibitem[{Gordon {et~al.}(2022)Gordon, Rothman, Hargreaves, Hashemi, Karlovets, Skinner, Conway, Hill, Kochanov, Tan, Wcisło, Finenko, Nelson, Bernath, Birk, Boudon, Campargue, Chance, Coustenis, Drouin, Flaud, Gamache, Hodges, Jacquemart, Mlawer, Nikitin, Perevalov, Rotger, Tennyson, Toon, Tran, Tyuterev, Adkins, Baker, Barbe, Canè, Császár, Dudaryonok, Egorov, Fleisher, Fleurbaey, Foltynowicz, Furtenbacher, Harrison, Hartmann, Horneman, Huang, Karman, Karns, Kassi, Kleiner, Kofman, Kwabia–Tchana, Lavrentieva, Lee, Long, Lukashevskaya, Lyulin, Makhnev, Matt, Massie, Melosso, Mikhailenko, Mondelain, Müller, Naumenko, Perrin, Polyansky, Raddaoui, Raston, Reed, Rey, Richard, Tóbiás, Sadiek, Schwenke, Starikova, Sung, Tamassia, Tashkun, {Vander Auwera}, Vasilenko, Vigasin, Villanueva, Vispoel, Wagner, Yachmenev, \& Yurchenko}]{hitran20}
Gordon, I., Rothman, L., Hargreaves, R., {et~al.} 2022, Journal of Quantitative Spectroscopy and Radiative Transfer, 277, 107949, \dodoi{https://doi.org/10.1016/j.jqsrt.2021.107949}

\bibitem[{H\"anni {et~al.}(2020)H\"anni, Altwegg, Pestoni, Rubin, Schroeder, Schuhmann, \& Wampfler}]{Hanni_Altwegg_Pestoni_Rubin_Schroeder_Schuhmann_Wampfler_2020}
H\"anni, N., Altwegg, K., Pestoni, B., {et~al.} 2020, Monthly Notices of the Royal Astronomical Society, 498, 2239–2248, \dodoi{10.1093/mnras/staa2387}

\bibitem[{H\"anni {et~al.}(2021)H\"anni, Altwegg, Balsiger, Combi, Fuselier, De~Keyser, Pestoni, Rubin, \& Wampfler}]{Hanni_Altwegg_Balsiger_Combi_Fuselier_De_Keyser_Pestoni_Rubin_Wampfler_2021}
H\"anni, N., Altwegg, K., Balsiger, H., {et~al.} 2021, Astronomy and Astrophysics, 647, A22, \dodoi{10.1051/0004-6361/202039580}

\bibitem[{Hardy {et~al.}(2023)Hardy, Richard, Boudon, Khan, Manceron, \& Dridi}]{HARDY2023108779}
Hardy, P., Richard, C., Boudon, V., {et~al.} 2023, Journal of Quantitative Spectroscopy and Radiative Transfer, 311, 108779, \dodoi{https://doi.org/10.1016/j.jqsrt.2023.108779}

\bibitem[{Haser(1957)}]{haser}
Haser, L. 1957, Bulletins de l'Acad{\'e}mie Royale de Belgique, 43, 740

\bibitem[{Herzberg(1950)}]{herzbergvol1}
Herzberg, G. 1950, Molecular Spectra and Molecular Structure: I. Spectra of Diatomic Molecules (New York: Van Nostrand Reinhold)

\bibitem[{Kunde {et~al.}(1981)Kunde, Aikin, Hanel, Jennings, Maguire, \& Samuelson}]{Kunde_1981}
Kunde, V.~G., Aikin, A.~C., Hanel, R.~A., {et~al.} 1981, Nature, 292, 686–688, \dodoi{10.1038/292686a0}

\bibitem[{Li {et~al.}(2025)Li, Shi, Ma, \& Sun}]{Li_2025}
Li, J., Shi, J., Ma, Y., \& Sun, J. 2025, The Astronomical Journal, 169, 126, \dodoi{10.3847/1538-3881/ada7e7}

\bibitem[{Liu \& Liu(2024)}]{liu}
Liu, B., \& Liu, X. 2024, A\&A, 683, A51, \dodoi{10.1051/0004-6361/202348663}

\bibitem[{Maki(2011)}]{MAKI2011166}
Maki, A.~G. 2011, Journal of Molecular Spectroscopy, 269, 166, \dodoi{https://doi.org/10.1016/j.jms.2011.06.002}

\bibitem[{{Milam} {et~al.}(2023){Milam}, {Roth}, {Villanueva}, {Wong}, {Kelley}, {Bockelee-Morvan}, \& {Hammel}}]{milam}
{Milam}, S.~N., {Roth}, N.~X., {Villanueva}, G.~L., {et~al.} 2023, in LPI Contributions, Vol. 2851, Asteroids, Comets, Meteors Conference, 2163

\bibitem[{Nuth \& Glicker(1982)}]{nuth}
Nuth, J.~A., \& Glicker, S. 1982, Journal of Quantitative Spectroscopy and Radiative Transfer, 28, 223, \dodoi{https://doi.org/10.1016/0022-4073(82)90025-5}

\bibitem[{Rank {et~al.}(1965)Rank, Pierre, \& Wiggins}]{RANK1965418}
Rank, D., Pierre, A., \& Wiggins, T. 1965, Journal of Molecular Spectroscopy, 18, 418, \dodoi{https://doi.org/10.1016/0022-2852(65)90048-2}

\bibitem[{Rao(2012)}]{rao2012chapter3}
Rao, K.~N. 2012, Spectroscopy of the Earth's Atmosphere and Interstellar Medium (Academic Press), 159

\bibitem[{{Rothman} \& {Young}(1981)}]{rothman}
{Rothman}, L.~S., \& {Young}, L.~D.~G. 1981, Journal of Quantitative Spectroscopy and Radiative Transfer, 25, 505, \dodoi{10.1016/0022-4073(81)90026-1}

\bibitem[{Rubin {et~al.}(2019)Rubin, Altwegg, Balsiger, Berthelier, Combi, De~Keyser, Drozdovskaya, Fiethe, Fuselier, Gasc, Gombosi, Hänni, Hansen, Mall, Rème, Schroeder, Schuhmann, Sémon, Waite, Wampfler, \& Wurz}]{rubin2019}
Rubin, M., Altwegg, K., Balsiger, H., {et~al.} 2019, Monthly Notices of the Royal Astronomical Society, 489, 594, \dodoi{10.1093/mnras/stz2086}

\bibitem[{Russo {et~al.}(2001)Russo, Mumma, DiSanti, Magee-Sauer, \& Novak}]{RUSSO2001162}
Russo, N.~D., Mumma, M.~J., DiSanti, M.~A., Magee-Sauer, K., \& Novak, R. 2001, Icarus, 153, 162, \dodoi{https://doi.org/10.1006/icar.2001.6678}

\bibitem[{Western(2017)}]{WESTERN2017221}
Western, C.~M. 2017, Journal of Quantitative Spectroscopy and Radiative Transfer, 186, 221, \dodoi{https://doi.org/10.1016/j.jqsrt.2016.04.010}

\end{thebibliography}

\end{document}